\documentclass[twocolumn]{aastex62}

\newcommand{\ctext}[1]{\raise0.2ex\hbox{\textcircled{\scriptsize{#1}}}}

\received{xxxx, 2021}
\revised{xxxx, 2021}
\accepted{Nov. 24, 2021}
\submitjournal{ApJ}

\shorttitle{WERGS : Distant Filamentary Structures Pointed by High-$z$ Radio Galaxies at $z\sim4$}
\shortauthors{Uchiyama et al.}

\begin{document}
\title{A Wide and Deep Exploration of Radio Galaxies with Subaru HSC (WERGS). VI. \\
 Distant Filamentary Structures Pointed by High-$z$ Radio Galaxies at $z\sim4$}
\correspondingauthor{Hisakazu Uchiyama}
\email{uchiyama@cosmos.phys.ehime-u.ac.jp}

\author{Hisakazu Uchiyama}
\affiliation{Research Center for Space and Cosmic Evolution, Ehime University, 2-5 Bunkyo-cho, Matsuyama, Ehime 790-8577, Japan} 

\author{Takuji Yamashita}
\affiliation{National Astronomical Observatory of Japan, Mitaka, Tokyo 181-8588, Japan} 

\author{Jun Toshikawa} 
\affiliation{Department of Physics, University of Bath, Claverton Down, Bath, BA2 7AY, UK}

\author{Nobunari Kashikawa}
\affiliation{Department of Astronomy, School of Science, The University of Tokyo, 7-3-1 Hongo, Bunkyo-ku, Tokyo, JAPAN, 113-0033}
\affiliation{Research Center for the Early Universe, The University of Tokyo, 7-3-1 Hongo, Bunkyo-ku, Tokyo 113-0033, Japan}

\author{Kohei Ichikawa} 
\affiliation{Astronomical Institute, Tohoku University, Aramaki, Aoba, Sendai 980-8578, Japan} 

\author{Mariko Kubo}
\affiliation{Research Center for Space and Cosmic Evolution, Ehime University, 2-5 Bunkyo-cho, Matsuyama, Ehime 790-8577, Japan} 

\author{Kei Ito}
\affiliation{Department of Astronomical Science, The Graduate University for Advanced Studies, SOKENDAI, Mitaka, Tokyo 181-8588, Japan}

\author{Nozomu Kawakatu}
\affiliation{National Institute of Technology, Kure College, 2-2-11, Agaminami, Kure, Hiroshima 737-8506, Japan}

\author{Tohru Nagao}
\affiliation{Research Center for Space and Cosmic Evolution, Ehime University, 2-5 Bunkyo-cho, Matsuyama, Ehime 790-8577, Japan}

\author[0000-0002-3531-7863]{Yoshiki Toba}
\affiliation{Research Center for Space and Cosmic Evolution, Ehime University, 2-5 Bunkyo-cho, Matsuyama, Ehime 790-8577, Japan} 
\affiliation{Department of Astronomy, Kyoto University, Kitashirakawa-Oiwake-cho, Sakyo-ku, Kyoto 606-8502, Japan}
\affiliation{Academia Sinica Institute of Astronomy and Astrophysics, 11F of Astronomy-Mathematics Building, AS/NTU, No.1, Section 4, Roosevelt Road, Taipei 10617, Taiwan}

\author{Yoshiaki Ono}
\affiliation{Institute for Cosmic Ray Research, The University of Tokyo, 5-1-5 Kashiwanoha, Kashiwa, Chiba 277-8582, Japan}

\author{Yuichi Harikane}
\affiliation{Institute for Cosmic Ray Research, The University of Tokyo, 5-1-5 Kashiwanoha, Kashiwa, Chiba 277-8582, Japan} 
\affiliation{Department of Physics and Astronomy, University College London, London, WC1E 6BT, UK} 

\author{Masatoshi Imanishi}
\affiliation{National Astronomical Observatory of Japan, Mitaka, Tokyo 181-8588, Japan} 

\author{Masaru Kajisawa}
\affiliation{Research Center for Space and Cosmic Evolution, Ehime University, 2-5 Bunkyo-cho, Matsuyama, Ehime 790-8577, Japan} 

\author{Chien-Hsiu Lee} 
\affiliation{NSF's National Optical-Infrared Astronomy Research Laboratory, Tucson, AZ 85742, USA} 

\author{Yongming Liang}
\affiliation{Department of Astronomical Science, The Graduate University for Advanced Studies, SOKENDAI, Mitaka, Tokyo 181-8588, Japan}

\begin{abstract} 
We present the environmental properties around high-$z$ radio galaxies (HzRGs) at $z\sim4$, which have been poorly investigated because of their  rarity. 
We use the largest samples of HzRGs and $g$-dropout galaxy overdense regions at $z\sim4$, which were constructed from Hyper Suprime-Cam Subaru Strategic Program, to characterize the HzRG environments statistically.   
We measure the $g$-dropout galaxy overdensities around 21 HzRGs  whose rest-frame 1.4 GHz radio luminosities ($L_{1.4\mathrm{GHz}}$) are  $10^{26-27}$ W Hz$^{-1}$. 
We find that the overdensities around the faint HzRGs with $L_{1.4\mathrm{GHz}}\sim10^{26.0-26.5}$ W Hz$^{-1}$ tend to be higher than that of the $g$-dropout galaxies. On the other hand, no significant difference of density environments is found between the luminous HzRGs  with $ L_{1.4\mathrm{GHz}}\sim10^{26.5-27.0} $ W Hz$^{-1}$ and the $g$-dropout galaxies.  
The HzRGs are found to occupy more massive halos than $g$-dropout galaxies through a cross-correlation between the HzRGs and $g$-dropout galaxies. 
This trend is more pronounced in the faint HzRGs. 
These results are consistent with a scenario where HzRGs get older and more massive as the radio-luminosity decreases.
The HzRGs are expected to trace the progenitors of local cluster halos from their calculated halo mass. 
In addition, we find that surrounding galaxies tend to distribute along the radio-jet major axis of the HzRGs at angular distances less than $\lesssim500$ physical kpc. 
Our findings imply the onset of the filamentary structures around the HzRGs at $z\sim4$. 
\end{abstract}

\keywords{Radio galaxies; Active galaxies; High-redshift galaxies; Lyman-break galaxies }

\section{INTRODUCTION}  
The formation and evolution of radio galaxies, which host super massive black holes (SMBHs) launching radio jets/lobes, have strong dependence on their surrounding environments \citep[e.g.,][]{Best05}. 
According to the Blandford-Znajek process  \citep[][]{Blandford77}, the power of a radio jet/lobe monotonically increases with the black hole mass and spin. 
These physical quantities are built up by gas accretion and/or SMBH mergers, which are mostly induced by galaxy mergers \citep[e.g.,][]{Fanidakis11}.  
Thus,  radio galaxies are believed to reside in the galaxy overdense regions (massive halos) where galaxy mergers are expected to occur more often \citep[e.g.,][]{Hatch14,Chiaberge15,Overzier16}. 
On the other hand, radio galaxy is an energy dissipation system; the propagation of radio jet/lobe can extend beyond the host galaxy halo \citep[e.g.,][]{Jamrozy14, Lan20}.  
It can mechanically blow off substantial baryons and suppress star formation in the host and/or surrounding halos \citep[][]{Morganti05, McNamara07, Birzan08, Shabala11, Fabian12, Yuan16,  Nesvadba17, Izquierdo18}. 
These large-scale negative feedbacks could disturb such close relations between radio galaxies and its environment. 
Therefore, characterizing the environments of radio galaxies as a function of its properties at different redshifts is key to reveal the underlying baryonic physics in the co-evolution of radio galaxy and surrounding galaxies.

There have been considerable studies to examine radio galaxy environments by direct imaging/spectroscopic observations for their surrounding galaxies at $z<3$ \citep[For details, see the comprehensive review by][]{Overzier16}. 
Most of the radio galaxies tend to exist in the overdense regions \citep[e.g.,][]{Gendre13, Ramos13,  Hatch14,  Karouzos14,  Pace14, Jones15, Malavasi15, Croston18, 
Magliocchetti18, Kolwa19, Massaro19, Moravec19, Penney19, Shen19}. 
Additionally, the dark matter halo mass of radio galaxies are estimated to be $\gtrsim10^{13} M_\odot$ \citep[][]{Peacock91,Mag04,Brand05,Lindsay14,Mag16,Lindsay14b}, implying that radio galaxies could be associated with cluster progenitors. 

Clustering strength of local radio galaxies tends to depend on their radio luminosities; 
the sign of the correlation coefficient can be positive or negative according to the luminosity range \citep[e.g.,][]{Donoso10, Falder10, Husband16}.  
In some cases, a dependency of the clustering strength of radio galaxies on the radio luminosities can be explained by the abundance ratio of high-excitation radio galaxies (HERGs)/low-excitation radio galaxies (LERGs) in radio galaxies \citep[][]{Donoso10}. 
Several studies 
showed that  LERGs tend to reside in higher density regions than HERGs \citep[e.g.,][]{Ramos13, Ching17}. 
HERGs have radiatively-efficient accretion disks dominated by cold (quasar) mode accretion \citep[][]{Bower06, Croton06}. 
In this accretion channel, due to the strong radio jets, the cooling of the hot gas would be suppressed \citep[e.g.,][]{Ichikawa17}. 
Hence, the formation of magnetically choked accretion flow (MCAF) is suggested to be caused through the cold mode accretion followed by hot mode accretion \citep[][]{Sikora13}. 
This channel occurs dominantly in the dark matter halos with the mass of $\sim10^{12} h^{-1} M_\odot$ \citep[][]{Orsi16}. 
LERGs, on the other hand, are hosted by more massive halos with hot-halo mode accretion \citep[][]{Turner15, Orsi16}. 
In this massive system, it may be more difficult for major mergers or cold stream to occur \citep[][]{Keres05, Dekel06}. 
The HERG and LERG are essentially different objects that cannot be unified in the active galactic nuclei (AGN) unification model; 
HERGs are unified with radio-loud quasars (RLQSOs), broad-line radio galaxies, and narrow-line radio galaxies, while LERGs are identical with BL Lac objects with no conspicuous emission lines \citep[e.g.,][]{Ching17}. 

Even beyond $z\sim4$, galaxies are  found to be strongly clustered around high-$z$ radio galaxies, ``HzRGs" \citep[][]{Overzier06, Venemans07, Kuiper11}.  
HzRGs are considered to be progenitors of the brightest cluster galaxies \citep[``proto-BCGs"; e.g., ][]{Ito19, West94}. 
On the other hand, \citet{Kikuta17} and \citet{Zeballos18} found that galaxies are not densely distributed around the HzRGs at $z\sim4$. 
So far,  the number of HzRGs whose environments have been studied is critically small \citep[$\simeq3$ at $z\sim4$; see ][]{Chiang13, Overzier16}. 
Especially, the HzRGs are limited to be the brightest ones with the rest-frame 1.4 GHz luminosity, $L_{1.4\text{GHz}}$,  of $>10^{27}$ W Hz$^{-1}$, meaning lack of generality of the HzRG environments at the epoch. 
This is due to the deficiency of deep and wide radio/optical surveys, which do not allow the detection of rare  very bright HzRGs as well as faint ones, and sufficient surrounding galaxies to characterize their environments.   
The small sample size also limits the estimation of the mean halo masses of the HzRGs. 
Constructing a statistical sample of fainter HzRGs  
and examining a possible correlation between the properties of HzRGs and galaxy overdense regions are essential to obtain the general picture of the system.  


Investigating a relation between the jet orientation of HzRG and the surrounding galaxy spatial distribution is also crucial to give 
insight into the HzRG-galaxy co-evolution physics.  
Some studies show that the surrounding galaxies around radio galaxies are distributed along the radio major axis \citep[``companion alignment": e.g., ][]{Venemans07}. 
One of the possible causes of the alignment is the anisotropic merging of galaxies \citep[e.g.,][]{West94}. 
Galaxy mergers are thought to occur anisotropically following the surrounding large-scale environments \citep[e.g.,][]{West94}. 
Radial-orbit galaxy mergers along filamentary structure \citep[e.g.,][]{Gonzalez16} can lead to remnants with mass distributions similar to prolate spheroids \citep[e.g.,][]{Ebrova17, Li18, Drakos19}. 
The major axes of the spheroids tend to be aligned with the separation orientation of the two pre-merger objects \citep[e.g.,][]{West94, Kirk15}.     
The spin axes of the SMBHs in the remnants are expected to be aligned with the orientation of the angular momentum vector of the accreting material (that is, the major axes of the prolate spheroids) through the Lense-Thirring effect \citep[][]{Scheuer96, Natarajan98, King05}. 
Consequently, the radio jets launched from the SMBHs are expected to be oriented along the ridge of the galaxy overdense regions \citep[e.g.,][]{West94, Bornancini06}. 
In other words, galaxies are expected to be preferentially distributed along the radio jet orientation. 

Several cosmological $N$-body simulations, on the other hand, show the spin flip with respect to the halo masses: 
the spin of the low/high-mass halos tends to be parallel/perpendicular to the surrounding filamentary structures \citep[e.g.,][]{Aragon07, Zhang09, Codis12}. 
This spin flip could be explained by non-linear effect such as mergers or the constrained tidal torque theory \citep[e.g.,][]{Codis12, Codis15}. 
Based on these simulations, the jets launched from the most massive radio galaxies are likely to be anti-aligned with the ridge of the surrounding galaxy distribution.

The companion alignments around radio galaxies at $z\lesssim2$ are reported  \citep[][]{West94,Roettgering96,Pentericci00, Croft05, Bornancini06}. 
At higher-$z$, radio galaxies would be biased toward radio-luminous and massive ones due to observational limits. 
This would cause an anti-alignment of galaxies to the direction of the radio jets. 
In fact, \citet{Zeballos18} found that the surrounding massive sub-millimeter galaxies (SMGs) around radio galaxies tend to be anti-aligned with the jet axes.  
On the other hand, \citet{Stevens03} found the normal companion alignment signal of surrounding massive SMGs. 
\citet{Stevens03} sample had a small field of view, so they looked at companion galaxies close to the HzRGs. 
\citet{Zeballos18} looked at much wider maps, further away from the HzRGs. 
\citet{Venemans07} and \citet{Kikuta17} found no clear alignment feature for Lyman alpha emitters around radio galaxies. 
At $z\sim4$, the number of samples studied about companion alignment is $\sim3$. 
The small sample size at high-$z$ prevents us from capturing the general picture of companion alignment. 

Hyper Suprime-Cam \citep[HSC, ][]{Miyazaki12, Miyazaki18}, which is an unprecedented wide-field imaging instrument mounted on the 8.2 m Subaru telescope, allows us to build the largest sample of galaxy overdense regions and HzRGs beyond $z\sim4$. 
The Wide and Deep Exploration of RGs with Subaru HSC \citep[``WERGS", ][]{Yamashita18,Toba19, Yamashita20, Ichikawa21} is one of  the largest survey for radio galaxies using the extremely-wide imaging data produced by the HSC-SSP (Hyper Suprime-Cam Subaru Strategic Program), supplemented by the 1.4 GHz radio continuum catalog of the Faint Images of the Radio Sky at Twenty-cm \citep[``FIRST''; ][]{Becker95, Helfand15} survey. 
The Lyman break galaxies at $z\ge4$ extracted from HSC-SSP data \citep[][]{Ono18} are matched with the FIRST sources, leading to   the construction of a uniform HzRG sample at $z\ge4$ (Yamashita et al. 2021, in prep.). 
One HzRG at $z=4.72$ has been spectroscopically confirmed \citep[][]{Yamashita20a}. 
The $L_{1.4\text{GHz}}$ of the HzRGs at $z\sim4$ are estimated to be $\sim10^{26-27}$ W Hz$^{-1}$  (Yamashita et al. 2021, in prep.), which is $\gtrsim1$ dex fainter than what is mentioned in previous studies \citep[][]{Overzier06, Venemans07, Kuiper11}. 
The largest overdense region (protocluster candidate) sample at $z\sim4$ has been constructed, based on the surface number density measurement \citep[][]{Toshikawa18}. 
These effective methods have extended the sample size by a factor of $\gtrsim10$ for  both the HzRGs and the protocluster candidates at $z\sim4$. 

In this paper, we use both of the HzRG and galaxy overdense region samples based on the early data release of HSC-SSP to characterize the HzRG environments at $z\sim4$ statistically. 
Especially, we investigate whether the HzRGs are good tracer of protoclusters, and examine a possible  dependency of the overdensities around HzRGs on the radio luminosity or jet orientation.  
Moreover, we estimate the dark matter halo mass of the HzRGs from a cross-correlation  between the HzRGs and the $g$-dropout galaxies. 
This, for the first time, could give new insights and perspective on the general overview of HzRG environments at $z\sim4$. 

The paper is organized as follows. 
In Section 2, we describe the HSC-SSP survey and the construction of the overdense region and HzRG sample. 
In Section 3, we examine a possible correlation between the HzRGs and their surrounding densities. It includes the investigation on the galaxy density environment of each HzRG, whether the HzRGs reside in the galaxy overdense regions or not, the overdensity significance distribution around HzRGs as a function of their radio luminosities, and the relation between the surrounding galaxy distribution and the radio major axis of the HzRG. 
We also estimate the HzRG halo mass by clustering analysis. 
The implications of our results are discussed in Section 4. 
Finally, in Section 5 we conclude and summarize our findings. 
We assume the following cosmological parameters: 
$\Omega_{M} = 0.3 $, $\Omega_{\Lambda} = 0.7$, $H_{0} = 70 ~ $km s$^{-1}$ Mpc$^{-1}$ $=100 ~ h ~ $km s$^{-1}$ Mpc$^{-1}$, and magnitudes are given in the AB system. The normalization of the matter power spectrum, $\sigma_8$, is settled to be $0.84$.

\section{DATA AND SAMPLE SELECTION}
\subsection{Subaru HSC-SSP}
The Subaru HSC-SSP survey is an unprecedented deep-and-wide optical survey using HSC with 116 2K $\times$ 4K Hamamatsu fully-depleted CCDs and  a field-of-view of 1.$^\circ$5 diameter. 
In the present study,  we use the ``Wide layer" data of DR S16A \citep[][]{Aihara18a}, which consists of wide field images of $>200$ deg$^2$ with a median seeing of $0.\arcsec6 - 0.\arcsec8$ and taken by the five optical filters $g$, $r$, $i$, $z$, and $y$. 
The Wide layer data consists of five independent fields (i.e. W-XMMLSS, W-Wide12H, W-GAMA15H, W-HECTOMAP, and W-VVDS). 
The survey design and the filter information are given in \citet{Aihara18b} and \citet{Kawanomoto17}, respectively. 
\citet{Komiyama18} should be the reference for the camera system and the CCD dewar designs. 
Among the HSC-SSP observations, the on-site quality assurance system for the HSC \citep[OSQAH; ][]{Furusawa18} was used in order to provide a feed back to the running observation instantly. 
The dedicated pipeline hscPipe \citep[version 4.0.2, ][]{Bosch18}, which is a modified version of the Large Synoptic Survey Telescope software stack \citep[][]{Ivezic08, Axelrod10, Juric15}, was used for data reduction.  
The astrometric and photometric calibrations are associated with the Pan-STARRS1 system \citep[][]{Schlafly12, Tonry12, Magnier13}. 
To estimate fluxes and colors of sources, we use cModel magnitude, which is measured by fitting two components that are PSF-convolved galaxy models (de Vaucouleurs and exponential) to the source profile \citep[]{Abazajian04}.

\begin{figure}
\begin{center}
\includegraphics[width=1.0\linewidth]{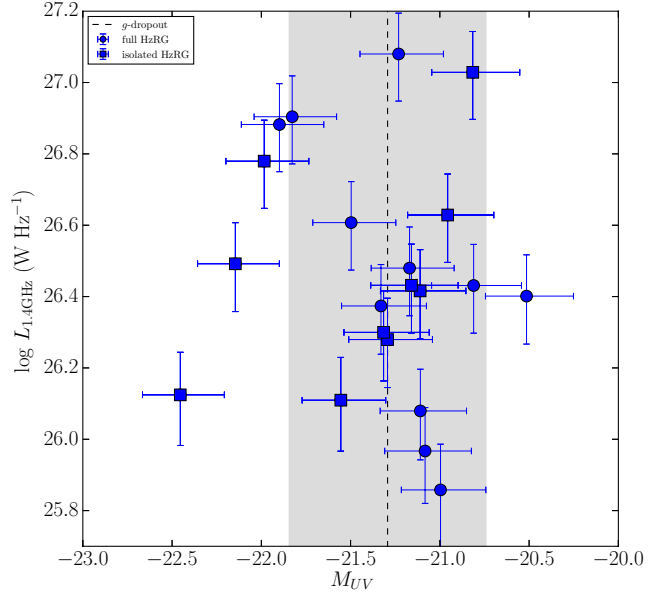}
\end{center}
\caption{Relation between radio and UV luminosities of HzRGs.  
The blue symbols show full HzRG sample. 
The isolated HzRGs are shown by the squares. 
The dashed line and gray shaded region show the average and standard deviation of the $g$-dropout galaxies with $i<25$. 
 }\label{uvradio}
\end{figure}

\begin{figure*}
\begin{center}
\includegraphics[width=0.9\linewidth]{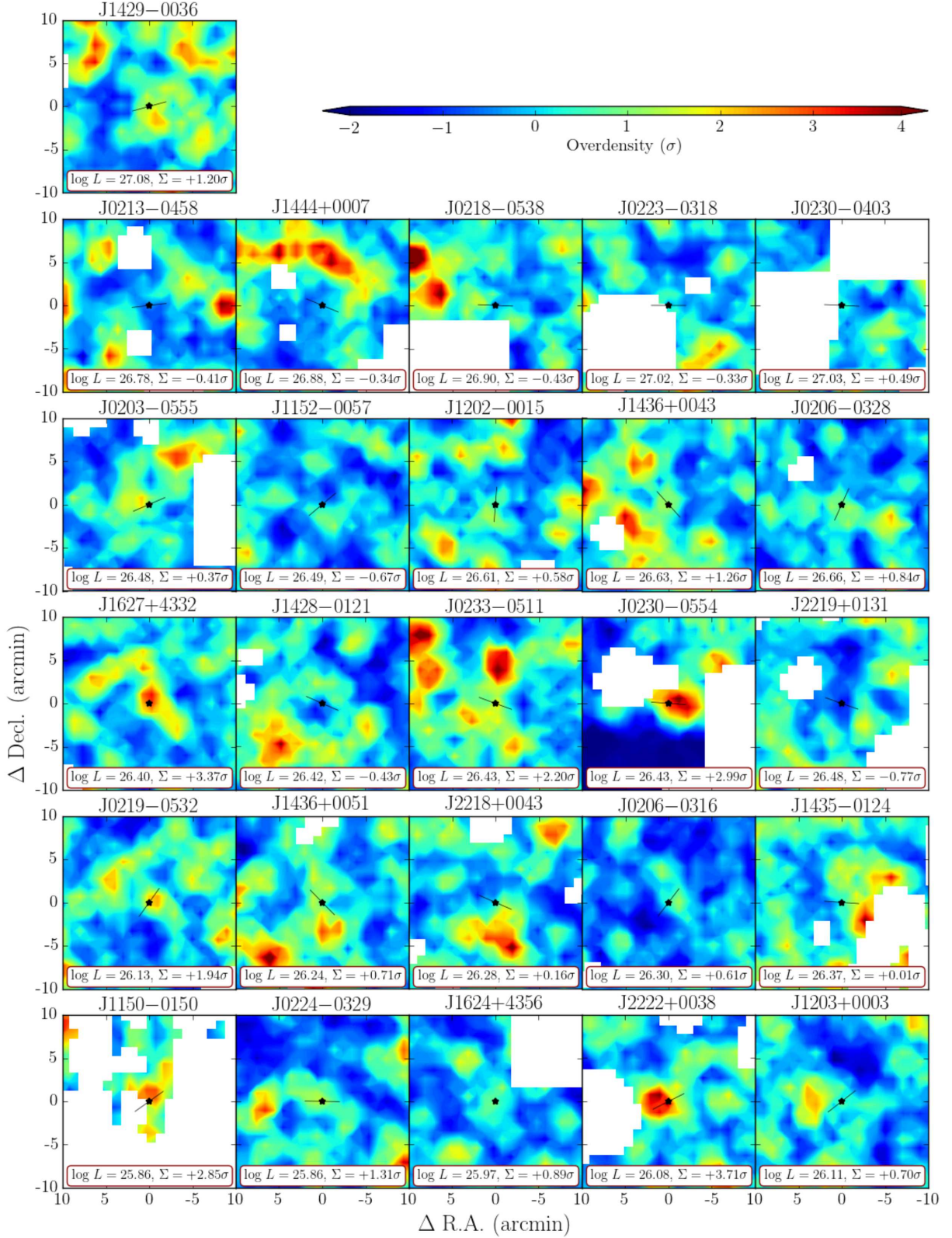}
\end{center}
\caption{
Overdensity map around each HzRG and RLQSO. 
Color contours show the overdensity significance, black stars indicate the HzRGs/RLQSOs positions, and the masked regions are shown in white. 
Black lines show the position angle of the radio morphology for each HzRG/RLQSO.  
The spatial resolution of the overdensity significance map is $1\times1$ arcmin$^2$. 
Both the HzRG/RLQSO 1.4 GHz luminosity and the overdensity significance at the grid point closest to the HzRG/RLQSO are indicated at the bottom of each panel.
 }  \label{rg}
\end{figure*}

\subsection{ Definition of overdense regions and protocluster candidates}
We use the catalogs of protocluster candidates and overdensity maps at $z\sim4$ that are described in \citet{Toshikawa18}. 
We enumerate the key steps of their construction below. 

First, $z\sim3.8$ galaxy candidates called $g$-dropout galaxies are selected by using the Lyman break technique.  
The $g$-dropout galaxies are selected from the five independent fields of the Wide layer data, using the following criteria: \citep[][]{vanderBurg10}. 
\begin{eqnarray}
  1.0  & < & g-r,  \label{eq1}\\ 
          r-i &  < &1.0, \\
1.5 (r-i) & < &(g-r) -0.80, \\ 
 r  & \le & m_{\mathrm{lim}, 3\sigma}, \\ 
  i  & \le & m_{\mathrm{lim}, 5\sigma},  \label{eq5}
\end{eqnarray}
where $m_{\mathrm{lim}, 3\sigma}$ and $m_{\mathrm{lim}, 5\sigma}$ are the limiting magnitude of $3\sigma$ and $5\sigma$, respectively. 
If the objects are not detected in the $g$-band filter at 3$\sigma$, their $g$-band magnitudes are  replaced by the corresponding 3$\sigma$ limiting magnitudes. 
All but homogeneous depth regions, where the $5\sigma$ limiting magnitudes in $g$-, $r$-, and $i$-bands are $>26.0$, $>25.5$, and $>25.5$, respectively, are masked. 
The regions around bright objects are also masked \citep[][]{Coupon18}. 
As a result, the total effective area used for this study is $121$ square degrees. 
In this area, 259,755 $g$-dropout galaxies down to $i<25.0$ are obtained.  
However, red galaxies at intermediate redshifts and dwarf stars could satisfy the $g$-dropout color selection criteria. 
We find that the rate of contamination from these objects is $\sim5$ \%  at $i<25.0$, which is estimated by using the same method\footnote{ 
Sources that do not satisfy the $g$-dropout selection criteria are selected from the UD-COSMOS catalogue in the Ultra-deep layer of the HSC-SSP. 
Among them, those that satisfy the $g$-dropout selection criteria in the Wide-layer depth are considered as foreground interlopers.
Then, we can estimate the contamination rate by comparing the surface densities of those interlopers and $g$-dropout galaxies. 
} as \citet[]{Ono18}. 
The number count is consistent with that of \citet{vanderBurg10}. 
The average redshift of the $g$-dropout galaxies is $z=3.8_{-0.5}^{+0.5}$ whose error is the full width at half maximum of the $g$-dropout selection function \citep[][]{Ono18, Toshikawa18}. 
The ultra violet (UV) absolute magnitudes ($M_{\mathrm{UV}}$) of the $g$-dropout galaxies are derived from the following relation given by \citet{Ono18}: 
\begin{equation}
M_{\mathrm{UV}} = i + 2.5 \log (1+z) - 5 \log \left( \frac{d_L(z)}{10 \mathrm{pc}} \right) + (m_{\mathrm{UV}} - i), \label{uv}
\end{equation}
where, $d_L(z)$ pc is the luminosity distance at $z$ and $m_{\mathrm{UV}}$ is the apparent magnitude at rest-frame UV wavelength. 
The term, $(m_{\mathrm{UV}} - i)$, represents the $K$ correction between the  $m_{\mathrm{UV}}$  and $i$. 
It is settled to be $0$ by assuming that the UV continua of $g$-dropout galaxies are flat \citep[][]{Ono18}. 

\begin{deluxetable}{lllllllll}[t!]
\tablecaption{$z\sim4$ HzRGs and RLQSOs \label{t1}}
\tablecolumns{14}
\tablenum{1}
\tablewidth{0pt}
\tablehead{
\colhead{ID \tablenotemark{a}} &  \colhead{log $L_{1.4\text{GHz}}$ \tablenotemark{b}} &  \colhead{$\Sigma$ \tablenotemark{c}}  & \colhead{P.A.  \tablenotemark{d}} \\
\colhead{} & \colhead{(W Hz$^{-1}$)} & \colhead{($\sigma$)} & \colhead{(deg)}}
\startdata
HSC J020343.79$-$055556.6 $**$  & $26.48_{-0.13}^{+0.12}$ & 0.37 & 113.4      \\ 
HSC J020612.23$-$031615.2 $\dagger$  & $26.30_{-0.14}^{+0.12}$ & 0.61 & 142.7      \\ 
HSC J020630.47$-$032847.1  $*$  & $26.66_{-0.13}^{+0.11}$ & 0.84 & 155.8        \\ 
HSC J021353.67$-$045818.0 $\dagger$  & $26.78_{-0.13}^{+0.11}$ & -0.41 & 96.7       \\ 
HSC J021803.39$-$053825.3  & $26.90_{-0.13}^{+0.11}$ &-0.43 & 87.7       \\ 
HSC J021952.75$-$053249.1 $\dagger$  & $26.12_{-0.14}^{+0.12}$ &1.94 & 144.4       \\ 
HSC J022320.70$-$031824.2  $*$  & $27.01_{-0.13}^{+0.11}$ & -0.33 & 89.9        \\ 
HSC J022454.95$-$032905.5 $**$  & $25.86_{-0.16}^{+0.13}$ &1.31 & 89.3       \\ 
HSC J023030.67$-$055440.8 $\dagger$  & $26.43_{-0.13}^{+0.12}$  & 2.99 & 85.7       \\ 
HSC J023032.41$-$040302.3 $\dagger$  & $27.03_{-0.13}^{+0.11}$ &0.49 & 87.4       \\ 
HSC J023303.17$-$051100.9   &  $26.43_{-0.13}^{+0.12}$ &2.20 & 69.8       \\ 
HSC J115038.16$-$015028.6   & $25.86_{-0.16}^{+0.13}$ & 2.85 & 125.5      \\ 
HSC J115208.95$-$005736.5 $\dagger$   & $26.49_{-0.13}^{+0.12}$ &-0.67 & 130.2      \\ 
HSC J120258.99$-$001543.8  & $26.61_{-0.13}^{+0.11}$ & 0.58& 175.8      \\ 
HSC J120343.34$+$000313.1 $\dagger$  & $26.11_{-0.14}^{+0.12}$ &0.70 & 128.0          \\ 
HSC J142815.62$-$002130.9  $\dagger$  & $26.42_{-0.13}^{+0.12}$  &-0.43 & 66.4     \\  	      
HSC J142949.68$-$003616.5  & $27.08_{-0.13}^{+0.11}$ &1.20 & 105.4     \\ 
HSC J143513.55$-$01244.8     & $26.37_{-0.14}^{+0.12}$ &0.01 & 85.8     \\ 
HSC J143618.45$+$004309.1 $\dagger$   & $26.63_{-0.13}^{+0.11}$ &1.26 & 41.9        \\ 
HSC J143634.50$+$005111.9 $*$  & $26.24_{-0.14}^{+0.12}$  &0.71 & 44.4      \\ 
HSC J144422.78$+$000700.0  & $26.88_{-0.13}^{+0.11}$ &-0.34 & 67.5        \\
HSC J162404.42$+$435624.2  & $25.97_{-0.15}^{+0.12}$ &0.89 & --         \\
HSC J162713.96$+$433258.0  & $26.40_{-0.13}^{+0.12}$ &3.37& --         \\
HSC J221828.11$+$004301.6 $\dagger$  & $26.28_{-0.13}^{+0.12}$ &0.16 & 66.5      \\
HSC J221959.44$+$013124.2  & $26.48_{-0.13}^{+0.12}$ & -0.77& 70.6      \\
HSC J222252.11$+$003806.1  & $26.08_{-0.14}^{+0.12}$ & 3.71 & 116.7      
\enddata
\tablenotetext{a}{Object ID. $\dagger$ symbol shows  isolated HzRG.  IDs that are not given a symbol represent not isolated HzRGs.  $*$ symbol shows spectroscopic confirmed RLQSO, and $**$ symbol shows RLQSO candidate.  }
\tablenotetext{b}{Logarithmic scale of the rest-frame  1.4 GHz radio luminosity. Here, it is assumed that each HzRG exists at $z=3.8_{-0.5}^{+0.5}$, and has a radio slope of $\alpha=-0.7$. 
The error is estimated by convolving the photometric error with the redshift error. }
\tablenotetext{c}{ Overdensity significance at the grid point closest to the HzRG.  }
\tablenotetext{d}{Position angle (degrees east of north) derived from the elliptical Gaussian model of the source. This has been de-convolved to remove the blurring effect caused by the elliptical Gaussian point-spread function. 
The symbol ``-" indicates that the P.A. is not measured correctly because the sizes of the major and minor axes determined before deconvolution are smaller than the beam size. 
This could be due to noise \citep[][]{White97}. 
}
\end{deluxetable}

Next, the fixed aperture method is applied to determine the surface density contour maps of $g$-dropout galaxies. 
Apertures with a radius of  $1.8$ arcmin, which corresponds to $0.75$ physical Mpc (pMpc) at $ z \sim 3.8$, are uniformly distributed in the sky area of the Wide layer images in a 1 arcmin grid spacing. 
This aperture size is comparable with the typical protocluster size at this epoch with a descendant halo mass of $\gtrsim 10^{14}~ M_{\odot}$ at $z=0$ \citep[]{Chiang13}.  
To determine the overdensity significance quantitatively, the local surface number density is estimated by counting the $g$-dropout galaxies with $i<25.0$ within the fixed aperture. 
When a 1.8-arcmin aperture falls on a masked region, the surface density is estimated by replacing the masked area with the average density. 
When more than $50$ \% of the area is masked in an aperture, the sky position is excluded in the protocluster search. 
Then, the overdensity significance, $\Sigma$, is defined by ($N$-$\bar{N}$)/$\sigma$, where $N$ is the number of the $g$-dropout galaxies in an aperture, 
and $\bar{N}$ and $\sigma$ are the average and standard deviation of $N$, respectively. 
We define the regions with $\Sigma > 1.5 \sigma$ 
as overdense regions.  

Finally, the protocluster candidates are defined by the overdense regions with prominently large significance of $\Sigma>4\sigma$. 
This definition allows us to obtain good protocluster candidates with low completeness ($\sim6$\%) but high purity ($>76\%$).   
Actually, \citet{Toshikawa16} showed that $ > 76 \%$ of $> 4\sigma$ overdense regions of $g$-dropouts are expected to evolve into dark matter halos with masses of $> 10^{14} M_{\odot}$ at $z=0$. 
Each $>4\sigma$ overdense region is carefully checked, and $22$ fake detections of mainly spiral arms of local galaxies are removed. 
As a result, 179 protocluster candidates are obtained in the Wide layer data, which have $z \sim 3.8$ and overdensity significances ranging from $4$ to $10\sigma$.

According to \citet{Toshikawa18}, as a result of the projection effect, the protocluster candidates are biased to the richest structure, expecting an average descendant halo mass of $\sim5\times10^{14} M_{\odot}$. 
Via clustering analysis, 
their mean halo mass at $z\sim4$ is estimated to be $2.3_{-0.5}^{+0.5}\times10^{13} h^{-1} M_\odot$, which is comparable to galaxy group halo mass \citep[][]{Toshikawa18}. 
It should be noted that the success rate of this technique has already been established ($\sim75$ \%) by the previous study of the $\!\sim\! 4$  deg$^2$ of the CFHTLS Deep Fields \citep[]{Toshikawa16}.



\begin{figure*}[!t]
\begin{center}
\includegraphics[width=1.0\linewidth]{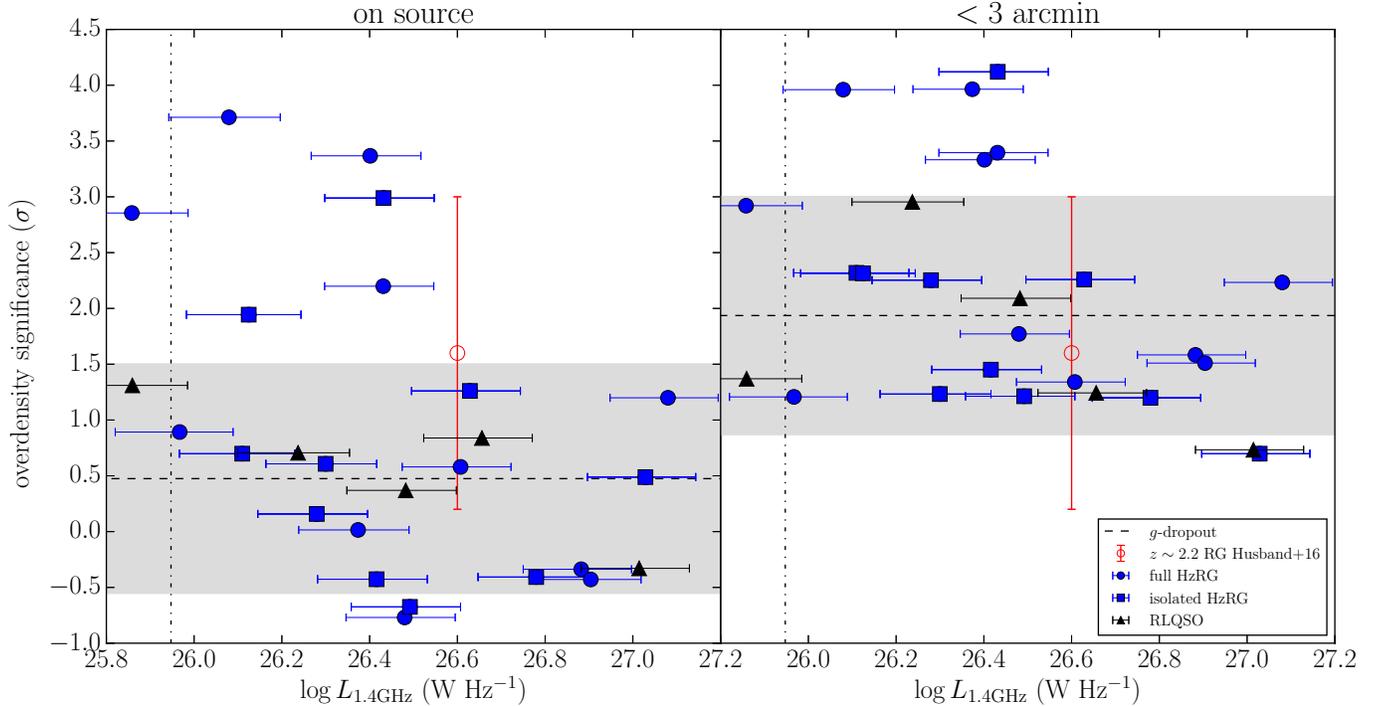}
\end{center} 
\caption{Dependency of the overdensities around the HzRGs on their rest-frame 1.4 GHz radio luminosities. In the left panel, the overdensity significances are estimated at the grid point closest to HzRGs/RLQSOs. 
The right panel shows the case of the local maximum overdensity significances within 3 arcmin radius centered on the HzRGs/RLQSOs.
The blue symbols show full HzRG sample, and the squares show the isolated ones. 
The RLQSOs are shown by the black triangles. 
The dashed line and the gray shaded region express the median 
and standard deviation of the overdensity significances of the $g$-dropout galaxies, respectively. 
The red open circle shows the case of a $z\sim2.2$ radio galaxy of \citet{Husband16} (see Section 4.2), estimated by the definition of the overdensity significance (Section 2.1) and $(\bar{N},\sigma)=(6.4,3.2)$ \citep[][]{Toshikawa18}.  
The vertical dashed-dotted line shows the rest-frame 1.4 GHz radio luminosity corresponding to the $1$ mJy flux density. 
The radio luminosities of the two sources are lower than the dashed-dotted line due to poor signal-to-noise ratio. 
}\label{sigrad}
\end{figure*}

\subsection{HzRGs} 
We use the $z\sim4$ HzRG sample constructed by Yamashita et al. (2021, in prep). 
We briefly summarize the essence of the HzRG selection as follows. 

First, Yamashita et al. extracted FIRST sources with  $>1$ mJy 1.4 GHz radio peak flux and the spurious detection probability of $< 0.05$ from the final release of the FIRST catalog \citep[][]{Helfand15}. 
The angular resolution of the FIRST survey is 5.4 arcsec. 
Then, they matched them with the $g$-dropout galaxy catalog of HSC-SSP \citep[][]{Ono18} in which the essentially same $g$-dropout selection criteria as those of the protocluster search are used, with a search radius of 1.0 arcsec, and obtained 35 matches in a 100 deg$^2$ field. 
Among the 35 matches, 5 are removed from the HzRG sample, because they are considered to be low-$z$ red galaxies by detections in SDSS $u$-band at $>2 \sigma$  significance ($< 23.1$ mag) and their optical appearances. 
Actually, two of the 5 low-$z$ red galaxies have archival spectroscopic redshifts of $\sim0.4$. 
Out of remaining 30 matches, 4 are spectroscopically confirmed high-$z$ quasars at $z\sim4$, namely radio-loud quasars (``RLQSOs").
These 4 RLQSOs are found in the Sloan Digital Sky Survey (SDSS) quasar catalog \citep[][]{Paris17}. 
Other two sources are also considered to be RLQSOs despite no spectroscopic confirmation, because these two source are stellar-like objects with $i < 24$ in the HSC $i$-band images \citep[][]{Akiyama18}. 
Thus, the two objects are good candidates of RLQSOs. 
Consequently, 24 matches are $g$-dropout HzRGs. 
In addition, 11 out of 24 HzRGs have no radio companions within a projected radius of 3 arcmin from each HzRG  (``isolated HzRGs"; Yamashita et al. in prep.). 
This isolated sample is constructed for the purpose of avoiding a possible case in which the position of a Lyman break galaxy coincides with  part of a radio sources at low-$z$ (e.g., widely separated radio lobes). 
Finally, we restrict the sample to sources that only reside in the effective area defined in the protocluster search. 
21 HzRGs (hereinafter called ``full HzRG sample"), 10 out of which are isolated HzRGs, and also 5 RLQSOs (= 2 RLQSO candidates $+$ 3 spectroscopic confirmed RLQSOs) are found in the effective area. 
This sample is listed in Table \ref{t1}. 

The rest-frame 1.4 GHz luminosities, $L_{1.4\text{GHz}}$, of the HzRGs and RLQSOs are calculated to be $\sim10^{25.8-27.2}$ W Hz$^{-1}$  
by using the total integrated radio flux density 
with an assumption of the radio spectrum index of $\alpha = -0.7$, which is a representative value of the parent sample of the HzRGs in this study (Yamashita et al., in prep.), and a redshift $z=3.8_{-0.5}^{+0.5}$ corresponding to the 
redshift of the $g$-dropout galaxies.  
These radio luminosities are much higher than $10^{23.5}$ W Hz$^{-1}$, corresponding to the dichotomy  between radioactive AGNs and star forming galaxies at $z\sim3.8$ \citep[e.g.,][]{Magliocchetti02, Mauch07}. 
Thus, the radio emission from the HzRGs is expected to be of AGN origin. 
The position angles (degrees east of north) of the FIRST emission were derived 
 by fitting an elliptical Gaussian model to the sources \citep[][]{Helfand15}. 
Here, the position angles are estimated 
after deconvolution to remove the blurring effect of the elliptical Gaussian PSF.
These quantities are also summarized in Table \ref{t1}. 
The $M_{\mathrm{UV}}$ of the HzRGs were also estimated through equation (\ref{uv}), assuming that the $K$ correction term is zero. 
Figure \ref{uvradio} shows a relation between the radio and UV luminosities of the HzRGs. 
The HzRG $M_{\mathrm{UV}}$ values are almost consistent with those of the $g$-dropout galaxies with $i < 25$. 
In addition, we find no dependency of the HzRG $M_{\mathrm{UV}}$ values on their radio luminosities.
The overdensity significance map around each HzRG/RLQSO is shown in Figure \ref{rg}. 
The J1150$-$0150 field is heavily masked,  
as it resides near the survey boundary, where limiting magnitudes are shallow. 

\section{ANALYSIS AND RESULTS} 
\subsection{Spatial correlation between HzRGs and protocluster candidates}
In order to examine the spatial correlation between the HzRGs and the protocluster candidates, we estimate the minimum projected distances from each HzRG to the peaks of the overdensity significance maps in the protocluster candidate regions \citep[][]{Uchiyama18}. 
As a result, 1 out of 21 HzRGs, HSC J0230$-$0554, is found to spatially associate with the protocluster candidates within a $1.8$ arcmin angular separation that corresponds to the radial size of the typical protocluster at $z\sim4$ \citep[][]{Chiang13}. 
The peak overdensity significance of the protocluster candidate region is $4.1 \sigma$. 
The proportion of the HzRGs associated with the protocluster candidates is $4.8_{-3.8}^{+5.0}$ \% whose error is assumed to be a binomial distribution \citep[][]{Gehrels86}. 
The typical protocluster radial size could extend to at most $\sim3.0$ arcmin at $z\sim4$  \citep[][]{Chiang13}. 
The number of protocluster candidates  
found within the 3.0 arcmin angular radius centered on the HzRGs is confirmed to be still one.

We also randomly select the same number of $g$-dropout galaxies as HzRGs, 
and estimate the proportion of these $g$-dropout galaxies associated with the protocluster candidates within the 1.8 and 3.0 arcmin separations \citep[][]{Uchiyama20}. 
We repeat this operation 10,000 times and evaluate the mean and standard deviation of the proportions. 
As a result, the proportion of the randomly-selected $g$-dropout galaxies associated with the protocluster candidates within the 1.8 (3.0) arcmin separation is estimated to be $0.6_{-0.6}^{+1.7}$ \% ($4.5_{-4.5}^{+4.5}$ \%). 
Thus,  the proportion of the HzRGs associated with the protocluster candidates is  comparable to that of the $g$-dropout galaxies in both cases of the separations (i.e., 1.8 and 3.0 arcmin) within $1\sigma$ error. 
This result suggests that HzRGs do not preferentially appear in  protocluster regions at $z\sim4$, compared to $g$-dropout galaxies. 

Even if the HzRGs are limited to the isolated ones, the result does not change. 
The proportions for all the samples are summarized in Table \ref{t3}. 
Note that protocluster candidates are selected by surface overdensity of $g$-dropout galaxies, which results in biased and/or incomplete sample of protoclusters. 
The completeness of our protocluster search based on $g$-dropout galaxies is expected to be less than 10\% (Section 2.1).

\begin{deluxetable*}{rrrrr}[t!]
\tablecaption{Proportion of the HzRGs/$g$-dropout galaxies associated with protocluster candidates.\label{t3}}
\tablecolumns{4}
\tablenum{2}
\tablewidth{0pt}
\tablehead{
\colhead{radius \tablenotemark{a}} & \colhead{full HzRG \tablenotemark{b}} & \colhead{$g$-dropout for full HzRG \tablenotemark{c}}& \colhead{isolated HzRG \tablenotemark{d}}  & \colhead{$g$-dropout for isolated HzRG \tablenotemark{e}} \\ 
\colhead{(arcmin)} & \colhead{(\%)} &\colhead{(\%)} & \colhead{(\%)} & \colhead{(\%)} }
\startdata 
$<1.8$   & $4.8_{-3.8}^{+5.0}$  & $0.6_{-0.6}^{+1.7}$   & $10.0_{-8.0}^{+10.7}$   & $0.6_{-0.6}^{+2.4}$     \\ 
$<3.0$   & $4.8_{-3.8}^{+5.0}$  & $4.5_{-4.5}^{+4.5}$   & $10.0_{-8.0}^{+10.7}$   & $4.5_{-4.5}^{+5.7}$ 
\enddata
\tablenotetext{a}{Radius to search protocluster candidates from HzRGs/$g$-dropout galaxies.    }
\tablenotetext{b}{The proportion of full HzRG sample associated with protocluster candidates within a given radius.  }
\tablenotetext{c}{Among the randomly selected $g$-dropout galaxies whose number is the same ($=21$) as full HzRG sample, the proportion of the $g$-dropout galaxies associated with protocluster candidates within a given radius.  }
\tablenotetext{d}{The proportion of isolated HzRG sample associated with protocluster candidates within a given radius. }
\tablenotetext{e}{Among the randomly selected $g$-dropout galaxies whose number is the same ($=10$) as isolated HzRGs, the proportion of the $g$-dropout galaxies associated with protocluster candidates within the given radius. }
\end{deluxetable*}

\subsection{Radio luminosity dependence of overdensity significances around HzRGs} 
We measure 
the overdensity significances at the grid points, where the apertures in the fixed aperture method are placed (Section 2.2), closest to the HzRGs 
to examine the relation between the overdensity significances and their $L_{1.4\text{GHz}}$. 
In the left panel of Figure \ref{sigrad}, 
at a regime of a high luminosity of log $L_{1.4\text{GHz}}\gtrsim26.5$, 
the overdensity significance distribution of the HzRGs are almost comparable with that of the $g$-dropout galaxies. 
Especially, at this high-luminosity regime, no HzRGs reside in the denser regions compared to the $g$-dropout galaxies. 
On the other hand,  at low radio luminosity, log $L_{1.4\text{GHz}}\lesssim26.5$, some of the HzRGs reside in the overdense regions with $\Sigma\sim2\sigma-3\sigma$.  
The discussion about this result is given in Section 4.2.   
The overdensity significances around the RLQSOs are also measured, and found to be consistent with those around the $g$-dropout galaxies within $1\sigma$ error. 

We also examine a local maximum value of the overdensity significances around each HzRG/RLQSO with a search radius of 3 arcmin, and we use this value as their environmental measures.  
Even if we use the local maximum overdensity significances to examine the dependency of the number density environments on their radio luminosity, their trends are nearly unchanged (right panel of Figure \ref{sigrad}). 
The isolated HzRGs also have the same tendency as the full HzRG sample. 

\begin{figure*}[!t]
\begin{center}
\includegraphics[width=1.0\linewidth]{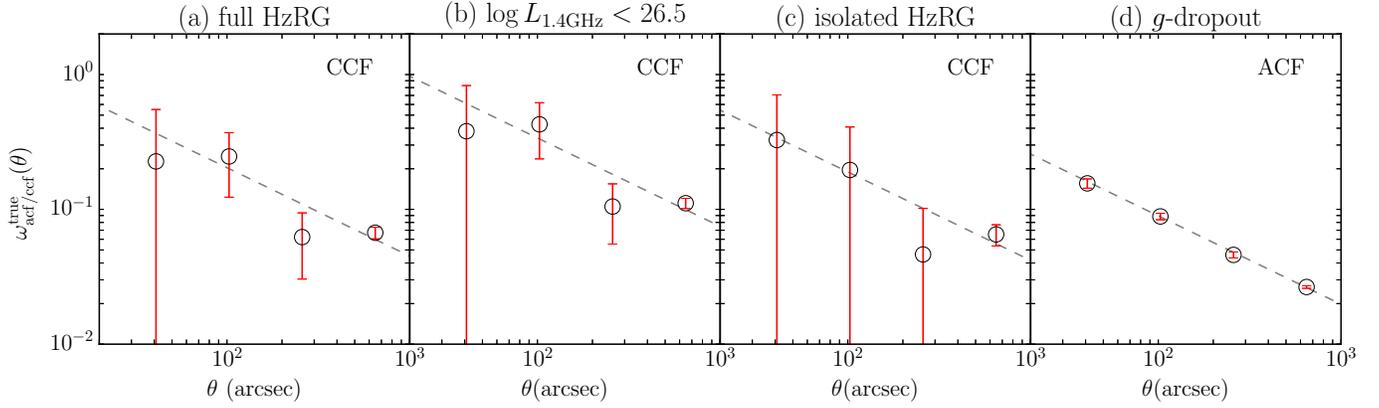}
\end{center} 
\caption{True cross-correlation function between $g$-dropout galaxies and (a) full HzRG sample, (b) the HzRGs with log $L_{1.4 \mathrm{GHz}} <26.5$, and (c) the isolated HzRGs. (d) true auto correlation function of the $g$-dropout galaxies.  The error bars are estimated with the Jackknife resampling. The dashed gray lines show the best-fit model functions, $\omega_{\mathrm{acf/ccf}}^{\mathrm{model}} (\cdot)$, provided in equation (\ref{model}).}\label{clustering}
\end{figure*}

\begin{figure*}[!t]
\begin{center}
\includegraphics[width=1.0\linewidth]{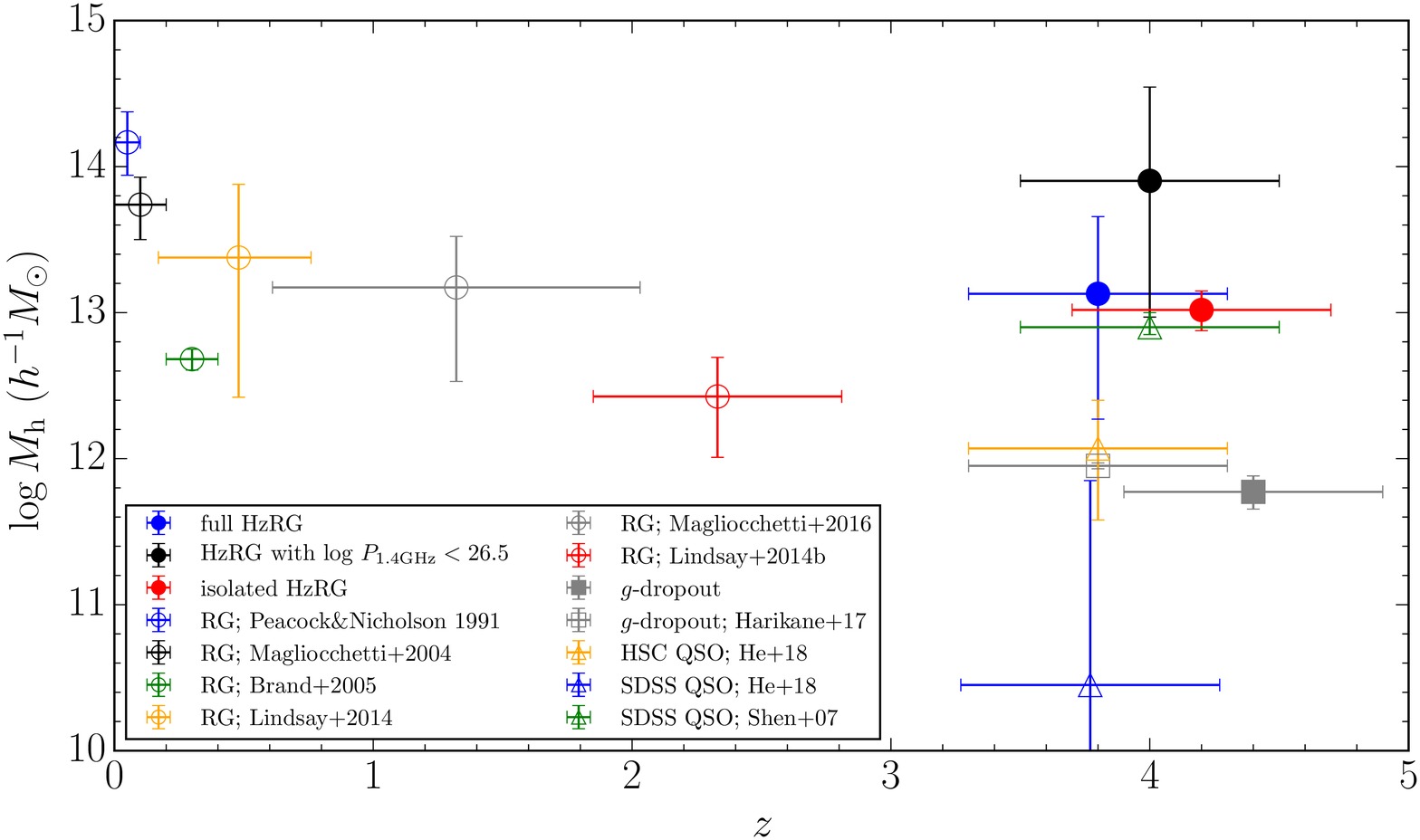}
\end{center} 
\caption{Halo masses of HzRGs, $g$-dropout galaxies, and quasars for a given redshift. The blue, black, and red filled circles, and gray filled square show the $z\sim4$ halo masses of full HzRG sample, the HzRGs with $\log L_{1.4 \mathrm{GHz}} <26.5$, the isolated HzRGs, and the $g$-dropout galaxies, respectively. For the sake of clarity in displaying the results at $z\sim4$, these values are shifted by $+0.2$ in the redshift direction. The halo masses of the $g$-dropout galaxies with $i<25.0$ estimated by \citet{Harikane17} is shown by the gray open square. The blue, black, orange, gray, and red open circles indicate the radio galaxy halo masses computed from the equations  (\ref{eq:corr} $-$ \ref{bb}), using the correlation lengths estimated by \citet{Peacock91}, \citet{Mag04}, \citet{Brand05}, \citet{Lindsay14}, \citet{Mag16}, and \citet{Lindsay14b}, respectively. The orange and blue open triangles show the quasar halo masses of HSC-detected and SDSS-detected quasars, estimated from a cross correlation, respectively \citep[][]{He18}. The SDSS-detected quasar halo mass at $z\sim4$ estimated by using an auto correlation function is also shown by a green open triangle \citep[][]{Shen07}. 
}\label{halomass}
\end{figure*}

\subsection{Halo Mass from Clustering Analysis}  
We estimate the halo masses of HzRGs from a cross-correlation between the full/isolated HzRG sample and $g$-dropout galaxies with $i<25.0$. 
We also examine a clustering strength of the low-luminosity HzRGs with log $L_{1.4 \mathrm{GHz}} < 26.5$, which tend to reside in the $>1.5 \sigma$ overdense regions (Section 3.2). 
The $g$-dropout galaxies and mask regions used in this clustering analysis are identical to  those used in the protocluster search. 
The details of the calculations are provided in Appendix A. 

Panel (d)  of Figure \ref{clustering} shows the true auto-correlation function (ACF), $\omega_{\mathrm{acf}}^{\mathrm{true}} (\cdot)$, of the $g$-dropout galaxies, given in equation (\ref{eq:10}). 
We fit the $\omega_{\mathrm{acf}}^{\mathrm{true}} (\cdot)$ with the model function, $\omega_{\mathrm{acf}}^{\mathrm{model}} (\theta) = A_\omega \theta^{-\beta}$, provided in equation (\ref{model}) through the $\chi^2$ minimization. 
As a result,  $A_\omega$ and $\beta$ are estimated to be $1.83_{-0.22}^{+0.22}$ and $0.66_{-0.03}^{+0.03}$, respectively. 
 Panels (a), (b) and (c) of Figure \ref{clustering} show the true cross-correlation function (CCF), $\omega_{\mathrm{ccf}}^{\mathrm{true}} (\cdot)$ of the full HzRG sample, the faint HzRGs with log $L_{1.4 \mathrm{GHz}} < 26.5$,  and the isolated HzRGs, respectively, given in equation (\ref{eq:10}). 
The uncertainties in $\omega_{\mathrm{ccf}}^{\mathrm{true}} (\cdot)$ are relatively large due to the small sample size. 
We also fit the $\omega_{\mathrm{ccf}}^{\mathrm{true}} (\cdot)$ to a model, $\omega_{\mathrm{ccf}}^{\mathrm{model}} (\theta) = A_\omega \theta^{-\beta}$, assuming that the index, $\beta$, of HzRGs is the same as that of the $g$-dropout galaxies ($\beta=0.66$) according to \citet{He18}. 
We find that $A_\omega$ are $4.2_{-1.7}^{+1.7}$ in the full HzRG sample,  $6.9_{-3.2}^{+3.2}$ in the HzRGs with  log $L_{1.4 \mathrm{GHz}} < 26.5$, and $3.8_{-0.2}^{+0.2}$ in the isolated ones.  
\footnote{We cannot estimate the $A_\omega$ for the luminous HzRGs with log $L_{1.4 \mathrm{GHz}} > 26.5$, due to their small sample size and weak clustering strength. }

Here, we focus only on the two-halo term of the ACF/CCF to examine the large-scale clustering. 
As the one-halo term of the $g$-dropout galaxies with $i<25.0$ can extend to $<20\arcsec$ at $z=3.8$ \citep[e.g.,][]{Ouchi05, Harikane17}, 
we used only a range of $\theta=20\arcsec-1000\arcsec$ to fit to the model, $\omega_{\mathrm{acf/ccf}}^{\mathrm{model}} (\cdot)$. 
In addition, the contamination correction of the ACF and CCF should be performed, but we do not apply the correction because the contamination rate is quite small ($\sim0.05$, see Section 2.1). 

\begin{deluxetable}{rrrrr}[t!]
\tablecaption{Result of clustering analysis. \label{t6}}
\tablecolumns{4}
\tablenum{3}
\tablewidth{0pt}
\tablehead{
\colhead{CF \tablenotemark{a}} & \colhead{$A_\omega$ \tablenotemark{b}} & \colhead{$r_0$ \tablenotemark{c}}& \colhead{$b_{\mathrm{hzrg/g}} $ \tablenotemark{d}}  & \colhead{log $M_h$ \tablenotemark{e}} \\ 
\colhead{}                                & \colhead{}                                                 & \colhead{$h^{-1}$ cMpc}            & \colhead{}                                                            & \colhead{$h^{-1} M_\odot$}       }
\startdata 
full HzRG CCF     & $4.2_{-1.7}^{+1.7}$  & $14.2_{-6.3}^{+6.5}$   & $10.6_{-4.5}^{+4.5}$  & $13.13_{-0.86}^{+0.53}$      \\ 
faint HzRG CCF  & $6.9_{-3.2}^{+3.2}$  & $24.5_{-10.5}^{+11.8}$   & $17.7_{-8.1}^{+8.1}$   & $13.90_{-0.93}^{+0.64}$  \\
isolated HzRG CCF & $3.8_{-0.2}^{+0.2}$ & $13.1_{-1.3}^{+1.3}$   & $9.9_{-0.9}^{+0.9}$   & $13.02_{-0.14}^{+0.13}$      \\ 
$g$-dropout ACF & $1.8_{-0.2}^{+0.2}$  & $6.2_{-0.5}^{+0.5}$   & $4.7_{-0.3}^{+0.3}$   & $11.77_{-0.12}^{+0.11}$     
\enddata
\tablenotetext{a}{Type of auto/cross correlation function.   }
\tablenotetext{b}{Amplitude of the CCF/ACF.  }
\tablenotetext{c}{Correlation length of each sample.  }
\tablenotetext{d}{Bias parameter of each sample.  }
\tablenotetext{e}{Typical halo mass of each sample.  }
\end{deluxetable}

We numerically solve the equations (\ref{eq:corr} $-$ \ref{bb}) to estimate the clustering length, bias, and typical halo mass of the HzRGs and $g$-dropout galaxies. 
The correlation length of the HzRGs is estimated by using the shooting method to their bias parameters numerically. 
The estimated values are summarized in Table \ref{t6}. 
Figure \ref{halomass} shows the radio galaxy halo masses as a function of redshift. 
The halo masses at $z<3$ are numerically computed from the equations  (\ref{eq:corr} $-$ \ref{bb}), using the correlation lengths estimated by previous studies \citep[][]{Peacock91,Mag04,Brand05,Lindsay14,Mag16,Lindsay14b}. 
We do not find a redshift-dependence of the clustering strengths of radio galaxies (open and filled circle symbols) from $z=1$ to $4$  
when considering the errors. 
This is consistent with the GALFORM simulation \citep[e.g.,][]{Orsi16} that shows the average halo masses of the radio galaxies at $z\sim2-4$ 
are  independent of redshift and approximately equal to $10^{13}M_\odot$. 
Less redshift-dependece  evolution of the radio galaxy clustering properties at $z<3$ has already been found by \citet{Mag16}. 
We also find that HzRGs tend to reside in more massive halos compared to the $g$-dropout galaxies. 
The difference in halo mass between the faint HzRGs with $\log L_{1.4 \mathrm{GHz}} <26.5$ and the $g$-dropout galaxies is found to be about six times larger with higher significance ($2.3\sigma$). 

It is worth to compare the estimated HzRG halo masses to those of the other AGN populations at $z\sim4$. 
\citet{He18} evaluated the quasar halo masses, using a cross correlation between the quasars and $g$-dropout galaxies. 
Their quasar sample is composed of two types; luminous ($M_{\text{UV}} =-29 \sim -26$) quasars detected by SDSS and less luminous ($M_{\text{UV}}=-26 \sim -22$) ones detected by HSC \citep[][]{Akiyama18}.  
We find that the halo masses of the HzRGs tend to be more massive than 
those of both quasar samples 
as shown in Figure \ref{halomass}.  
On the other hand, the SDSS quasar halo masses estimated through ACF \citep[][]{Shen07} is comparable to the HzRG ones. 
According to \citet{He18} and \citet{Uchiyama18}, the halo mass segregation between the SDSS quasar halo masses estimated from ACF and CCF  is thought to be due to the strong feedback from the luminous quasars \citep[``photoevaporation"; e.g.,][]{Uchiyama19}. 
This effect could yield the deficit of $g$-dropout galaxies in the vicinity of the quasars, leading to the lower amplitude of the cross correlation function.  
In fact, the SDSS-detected quasars tend to reside in halos with number densities similar to $g$-dropout galaxies   \citep[][]{Uchiyama18}. 
Our results suggest that HzRGs are better tracer of the galaxy overdense regions than quasars. 

\begin{figure*}
\begin{center}
\includegraphics[width=1.0\linewidth]{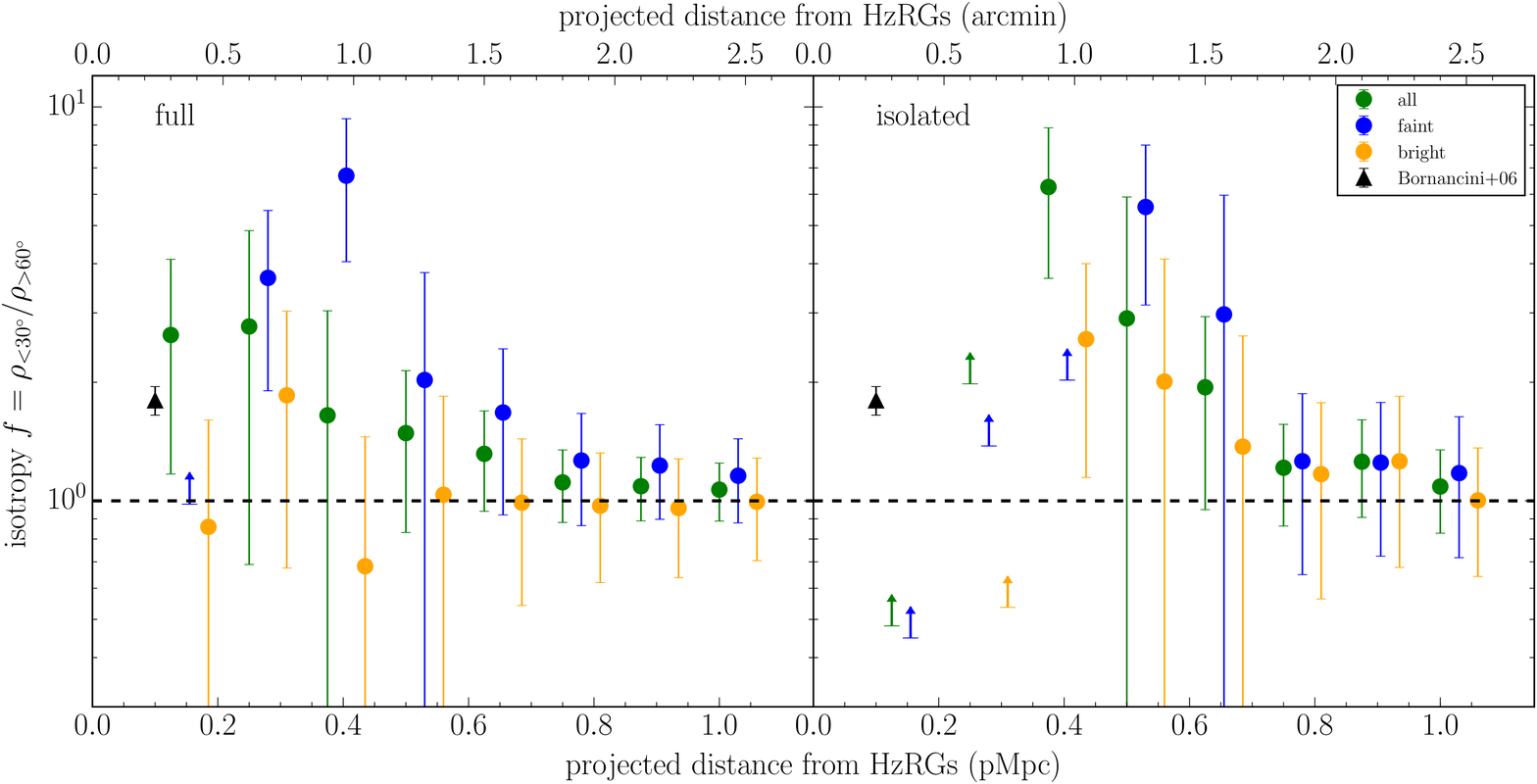}
\end{center}
\caption{Isotropy (green circles) as a function of the projected distance from full (left panel) and isolated (right panel) HzRG samples. 
The faint and luminous HzRGs are shown by the blue and orange circles shifted by $+0.03$ and $+0.06$ arcmin for the sake of the visibility, respectively. 
The error bar is the $1\sigma$ error estimated by the bootstrap method with 1,000 realizations. The black triangle shows the known isotropy estimated by \citet{Bornancini06} who use the $K$-band galaxies around 70 HzRGs at $z\sim0.1-3.8$ (see Section 4.3). The arrows show the lower limits of the isotropies of the HzRGs, given by \citet{Gehrels86}.
 }\label{isotropy}
\end{figure*}

\subsection{Companion alignment} 
To quantify the dependency of the galaxy distribution on the radio major axes of the HzRGs, we define the isotropy, $f(r)\equiv \rho_{<30}(r)/\rho_{>60}(r)$ \citep[e.g.,][]{Bornancini06}, where 
the $\rho_{<30}(r)$ and $\rho_{>60}(r)$ are the number densities of the $g$-dropout galaxies with $\Delta \phi <30$ degrees and $\Delta \phi >60$ degrees, respectively, in the effective area enclosed by a projected radius of $r$ centered on the HzRGs. 
 $\Delta \phi$ is the angle between the radio major axis of a HzRG and  the direction from the HzRG to each $g$-dropout galaxy. 
The spatial distributions of $g$-dropout galaxies around each HzRG are stacked into one, where each HzRG is centered and the radio major axes are aligned with each other. 
The isotropy of surrounding $g$-dropout galaxies is calculated in the stacked distribution as a function of a radius from the center.
We estimate $\Delta \phi$ for each $g$-dropout galaxy around the full and isolated HzRGs, and evaluate $\rho_{<30}(r)$ and $\rho_{>60}(r)$ by counting the number of the $g$-dropout galaxies lying in the given range of $\Delta \phi$ in their HzRG fields. 
The errors of the isotropy are measured by performing 1000 iterations of bootstrap resampling of the $g$-dropout galaxies. 
If $\rho_{>60}(r)=0$, that is, the number of the $g$-dropout galaxies with $\Delta \phi >60$ degrees, $N_{\Delta \phi >60}$, is zero, then we give the upper limit; $N_{\Delta \phi >60}=1.841$ corresponding to the $1\sigma$ Poisson single-sided upper limit  \citep[][]{Gehrels86}. 

Figure \ref{isotropy} shows  the isotropy as a function of the projected radii of $r$. 
We find that at $r\gtrsim1.5$ arcmin, the value of the isotropy is close to $1$ for  both of full and isolated samples, meaning that surrounding galaxies are randomly distributed far from the HzRGs.   
The isotropy gradually increases as the projected distance from the HzRGs decreases. 
The isotropy is  $2.6_{-1.5}^{+1.5}$ ($6.3_{-2.6}^{+2.6}$) at $r\sim0.3$ (0.9) arcmin for the full (isolated) sample.  
This means that in the vicinity of HzRGs, the rotational symmetry tends to be broken down in the projection space, and the $g$-dropout galaxies tend to distribute along their radio major axes.  
\citet{Bornancini06} examined the distribution of the $K$-band massive galaxies around 70 HzRGs at $z\sim0.1-3.8$ and found the isotropy $f=1.8_{-0.15}^{+0.15}$ at $\lesssim100$ pkpc, which is  consistent with our result within $1\sigma$ error as shown in Figure \ref{isotropy}. 

In order to examine if the distribution depends on the radio luminosities of the HzRGs, we divide the full/isolated HzRG samples into faint and luminous ones whose radio luminosities are fainter and brighter than the median  ($10^{26.5}$ erg s$^{-1}$) of the radio luminosities, respectively. 
Figure \ref{isotropy} also shows the isotropies of the faint and luminous HzRGs in both the full and isolated samples. 
We find that the isotropy is almost consistent with one across all distance scales for the luminous sample, while the isotropy estimated in the faint HzRG fields tends to be higher at $r\lesssim0.5$ pMpc in both of the full and isolated samples.   
The maximum values of the isotropy functions are $6.8_{-2.8}^{+2.3}$ and $5.5_{-2.4}^{+2.5}$ at $r\sim 1.0$ arcmin and $r\sim 1.3$ arcmin for the faint HzRGs of the full and isolated samples, respectively.  
If the $g$-dropout galaxies are uniformly distributed around the faint HzRGs, the isotropy is estimated as $1_{-0.51}^{+0.59}$ ($1_{-0.56}^{+0.65}$) at $r\sim 1.0$ (1.3) arcmin for the full (isolated) sample. 
Here, the error is assumed to follow a binomial distribution. 
Thus, the isotropy obtained from the observation is significantly higher, even if we take into account the error  from the uniform distribution. 
This means that the surrounding galaxies tend to distribute along the radio galaxies' major axes with faint HzRGs in their vicinities. 
The physical interpretation of this result is discussed in Section 4.3. 



\section{Discussion}

\subsection{Consistency with halo masses of HzRGs} 
We found that  $4.8_{-3.8}^{+5.0}$ \% ($10.0_{-8.0}^{+10.7}$ \%) of the full (isolated) HzRG sample could be associated with a protocluster candidate (Section 3.1). 
We examine if this result can be confirmed in terms of halo masses of HzRGs. 
The full and isolated HzRG samples have average halo masses, $1.3_{-1.2}^{+3.2} \times 10^{13}$ h$^{-1} M_\odot$ and  $1.0_{-0.3}^{+0.4} \times 10^{13}$ h$^{-1} M_\odot$, respectively (Section 3.4). 
The extended Press Schechter model \citep[][]{Press74, Bower91, Bond91, Lacey93} allows us to estimate the descendant halo masses of the HzRGs at $z=0$ \citep[][]{Uchiyama18, Uchiyama20}. 
By following the exact same procedure as \citet{Uchiyama20}, $94_{-90}^{+6}$ \%  ($85_{-34}^{+14}$ \%) halos of the full (isolated) HzRG sample at $z=3.8_{-0.5}^{+0.5}$ are found to evolve into the halos with mass of $>10^{14} M_{\odot}$ at $z=0$. 
Note that the completeness of the protocluster candidates we use is $\sim6$ \% (see Section 2.1). 
Since $\sim6$ \% of such halos that evolve into clusters at $z=0$ is expected to show an overdensity of $g$-dropout galaxies at $>4\sigma$  significance (Section 2.1), 
the expected proportion of the HzRGs associated with the protocluster candidates could be estimated to be $5.6_{-5.4}^{+0.4}$ \% ($5.1_{-2.0}^{+0.8}$ \%) for the full (isolated) sample. 
This proportion is consistent with our observational result, $4.8_{-3.8}^{+5.0}$ \% ($10.0_{-8.0}^{+10.7}$ \%) within $1\sigma$ error. 
This result means that HzRGs can trace the progenitors of cluster halos; but the protocluster search using surface number density is incomplete.  
Selection bias of the protocluster search may also prevent HzRGs from existing in $>4\sigma$ overdense regions of $g$-dropout star-forming  galaxies; 
quenching process might be starting to work for surrounding galaxies around HzRGs hosted by massive halos. 


\subsection{The HzRG environments as a function of radio luminosity} 
We found that only fainter HzRGs with log $L_{1.4\text{GHz}} < 26.5$  tend to reside in the galaxy denser regions / more massive halos compared to the $g$-dropout galaxies (Section 3.2 and Section 3.4). 
In this subsection, the implication and validity of this radio luminosity dependence are qualitatively discussed by comparing with previous studies at lower-$z$. 

\begin{figure}
\begin{center}
\includegraphics[width=1.0\linewidth]{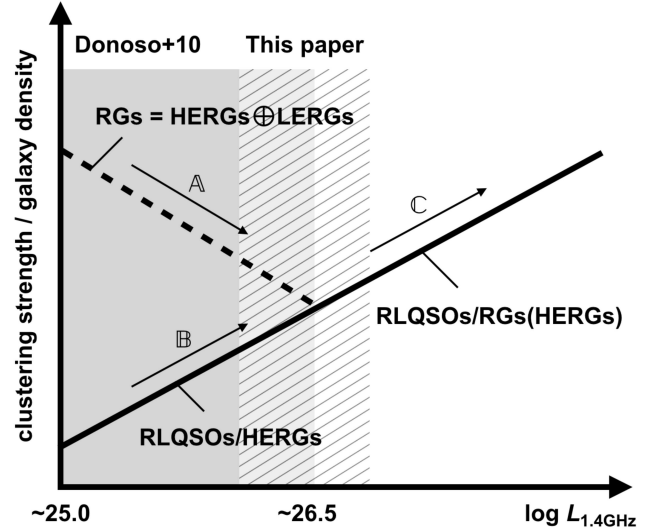}
\end{center}
\caption{
Sketch of the clustering strength of HzRG for given radio luminosity. The gray shade and hatched regions show the radio luminosity range studied by \citet{Donoso10} and this paper, respectively. The dashed and solid lines indicate the clustering strength of radio galaxies and RLQSOs, respectively. 
 }\label{punch}
\end{figure}

\citet{Donoso10} examined the clustering strengths of radio galaxies ($=$ HERGs $\oplus$ LERGs) and RLQSOs with log $L_{1.4\text{GHz}} \sim24.0-26.5$ at $z\sim0.4-0.8$. 
They found that the clustering strength of the radio galaxies is negative-correlated with the radio luminosity at log $L_{1.4\text{GHz}} \gtrsim 25.3$, as shown by arrow $\mathbb{A}$ in Figure \ref{punch}. 
They suggested that this negative correlation is caused by an increase in the proportion of HERGs as the radio luminosity increases \citep[e.g.,][]{Laing94, Jackson99, Best12}. According to \citet{Jackson99} and \citet{Best12}, the HERGs account for $\sim50$ \%  of radio galaxies at log $L_{1.4\text{GHz}} \sim26$, but $\sim100$ \% at log $L_{1.4\text{GHz}} \sim27$. 
HERGs tend to reside in lower dense regions (less massive halos) than LERGs \citep[e.g.,][]{Ching17}. 
In fact, \citet{Donoso10} confirmed that the clustering strength of the RLQSOs, which are unified with HERGs \citep[e.g.,][]{Urry95}, is relatively lower than that of the radio galaxies at log $L_{1.4\text{GHz}} \sim 25$ (Figure \ref{punch}). 
If their low-$z$ relations 
are assumed to also hold at high-$z$, the negative dependency of the local densities around our HzRGs (its luminosity range is shown with hatched region in Figure \ref{punch}) on the radio luminosity is expected to be caused by the change in the proportion of HERGs/LERGs among the HzRGs. 
In other words, most of the HzRGs would be classified into HERGs/RLQSOs at log $L_{1.4\text{GHz}} \sim 26.5-27.0$, while a non-negligible proportion of LERGs would populate in the HzRGs in the regime of log $L_{1.4\text{GHz}} \sim 26.0-26.5$. 
It is known that HERG host galaxies tend to have younger stellar populations with higher star formation rate than LERG in the low-$z$ universe \citep[][]{Smolcic09, Best12, Miraghaei17}.  
We find a consistent result in our sample; 
when examining the relation between the $M_{\mathrm{UV}}$ and on-source overdensity significances of our HzRGs, 
we find that they are positively correlated with $P$-value of 0.01 in Spearman-rank correlation test.   

Note that the negative correlation of HzRG clustering strength with the radio luminosity (arrow $\mathbb{A}$ in Figure \ref{punch}) is also consistent with a relation between the jet luminosity of radio galaxy and the ambient matter density which is expected to be higher in the galaxy overdense regions; 
if the matter density at the core radius ($\sim1$ kpc) of the host galaxy is high, the radio jets cannot spread beyond $\sim10$ kpc, resulting in FR I-like, that is, LERG-like radio galaxies with relatively low luminosity \citep[][]{Kawakatu09}.

The clustering strength of the RLQSOs/HERGs were also found to monotonically increase with the radio luminosity \citep[][]{Donoso10}, as shown by arrow $\mathbb{B}$ in Figure \ref{punch}. 
This could be explained by a scenario where the radio jet/lobe luminosities are thought to monotonically increase with the black hole mass and spin built up by galaxy mergers \citep[e.g.,][]{Fanidakis11} which preferentially occur in the galaxy overdense regions \citep[][]{Hatch14, Chiaberge15, Overzier16}, so that the radio luminosity could be positive-correlated with the overdensity (hereinafter, this relation is called ``$L_{\mathrm{jet}}-\Sigma$ relation"). 
\citet{Donoso10} suggested that the clustering strength of the radio galaxies should be consistent with that of RLQSOs at the high radio-luminous end. 
In fact, in our samples, we confirmed that at the radio luminous regime of log $L_{1.4\text{GHz}} \sim 26.5-27.0$, 
the HzRGs reside in the density environments similar to those of RLQSOs (Section 3.2). 

\begin{figure*}
\begin{center}
\includegraphics[width=1.0\linewidth]{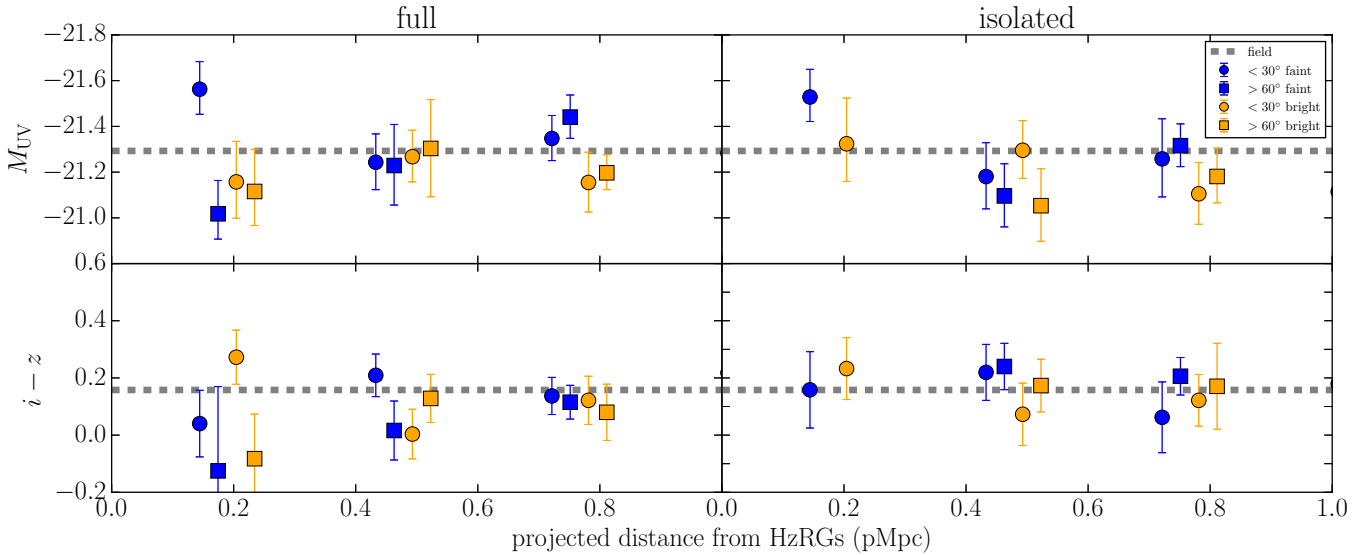}
\end{center}
\caption{
Average of the $M_{\mathrm{UV}}$ (upper panels) and $i-z$ colors (lower panels) of the $g$-dropout galaxies as a function of the projected distances from the faint (blue symbols) and luminous (orange symbols) HzRGs of the full (left panels) and isolated (right panels) samples. 
The $g$-dropout galaxies with $\Delta \phi <30$ degrees and  $\Delta \phi >60$ degrees are shown by circles and squares, respectively. 
Points in each bin are shifted by $+0.03$ arcmin in the horizontal axis for the sake of the visibility. 
The error bar is the standard error of the mean with the photometric and redshift errors of the objects convolved. 
The gray dashed lines indicate the average of the $g$-dropout galaxies in blank fields, respectively.  
The squares at $\sim0.2$ pMpc in right panels are not plotted, as no relevant $g$-dropout galaxies exist in this bin. 
 }\label{spalign}
\end{figure*}

We should examine whether our picture is consistent with the previous studies at a higher radio luminosity regime,  $\log L_{1.4\text{GHz}} > 27$. 
It is natural to assume that the unification between radio galaxies (HzRGs) and RLQSOs works at this regime. 
Actually, an observational evidence for the unification has been found in this luminosity regime \citep[][]{Wylezalek13}. 
Hyper-luminous radio galaxies/RLQSOs with $\log L_{1.4\text{GHz}} > 27$ are expected to reside in the galaxy denser regions according to the $L_{\mathrm{jet}}-\Sigma$ relation\footnote{The $L_{\mathrm{jet}}-\Sigma$ relation would break down in the extremely high-density regions as a very large velocity dispersion of galaxies in the massive structures reduces the merger frequency, and thus, SMBH spin and mass growth would be slow.    } as shown by arrow $\mathbb{C}$ in Figure \ref{punch}. 
There is definitive evidence that hyper luminous radio galaxies exist in the rich galaxy overdense regions \citep[][]{Venemans07, Falder10, Wylezalek13, Hatch14, Husband16}. 
For instance, \citet{Venemans07} examined the environments  of eight radio galaxy with log $L_{1.4\text{GHz}} \ge 27.3$ at $z\sim2-5$ and found   six out of them in overdense regions. 
These radio galaxies with available optical spectra are classified as either HERGs, RLQSOs, or dust-obscured RLQSOs \citep[][]{Eales93, Venemans07, Drouart16, Nesvadba17}. 
\citet{Hatch14} also found that radio-loud AGNs with log $L_{1.4\text{GHz}} > 27.2$ are significantly more present in high-density regions than radio-quiet AGNs at $z\sim1-3$. 
\citet{Falder10} found that there is a positive correlation between the radio luminosities and the ambient environments of HzRGs and RLQSOs with  log $L_{1.4\text{GHz}} = 26.65-28.65$ at $z\sim1$. 
The similar trend is also found for 7 luminous radio galaxies at $z\sim2.2$ \citep[][]{Husband16}. 
The faintest radio galaxy with log $L_{1.4\text{GHz}} = 26.6$   in \citet{Husband16}  resides in the lowest density region which is comparable to that of  blank fields (Figure \ref{sigrad}), and the others with log $L_{1.4\text{GHz}} > 27$  reside in the denser regions. 
On the other hand, \citet{Zeballos18} found that some of the HzRGs with log $L_{1.4\text{GHz}} > 27$ do not reside in the overdense regions of SMGs. 
However,  \citet{Zeballos18} could not deny the possibility that the overdense regions around those radio galaxies are buried due to the sample variance.
In future, we could assess hyper luminous radio galaxy environments at $z\sim4$; 
as the survey area of the HSC-SSP will achieve to $\sim1000$ deg$^2$, about 10 hyper luminous radio galaxies are expected to be found, assuming the radio galaxy luminosity function of \citet{Dunlop90}.  

The HzRGs that have fainter radio luminosities than those of our sample are also expected to reside in high-density regions (inverse of arrow $\mathbb{A}$ in Figure \ref{punch}). 
In fact, the low luminosity radio galaxies with log $L_{1.4\text{GHz}} \lesssim 25.8 $ W Hz$^{-1}$ are strongly clustered \citep[][]{Auger08, Donoso10,  Mawatari12, Karouzos14, Kolwa19}. 
Deeper radio observations could shed light on the HzRG environments in the fainter radio-regime. 


So far, our discussion has been based on the assumption that the HzRG properties such as clustering strength and proportion of HERGs and LERGs  do not change significantly with respect to redshift. 
\citet{Williams18} reported that the proportion of LERGs in a HERG $\oplus$ LERG sample decrease with redshift at $z=0.5-2.0$. 
On the other hand, we found that the HzRG halo mass are almost independent of redshift at $z\sim0.5-4.0$ (Section 3.4). 
Thus, the effect of redshift on reducing LERGs might not be large enough to change the negative correlation feature (arrow $\mathbb{A}$ in Figure \ref{punch}).
We need to statistically characterize HzRG environments on different redshift slices, $z<4$ or $z>4$ in order to confirm the redshift independency. 
Spectroscopic observations are also needed as our analysis is only based on photometric HzRG sample. 

\subsection{Radio jets point out large-scale structures}
We detected a companion alignment feature of $g$-dropout galaxies around faint HzRGs (Section 3.3). 
This result means that the filamentary structures are formed around HzRGs \citep[e.g.,][]{West94}. 
Such high-$z$ structures are expected to induce high star formation activity  \citep[e.g.,][]{Hayashi10}. 
On the other hand, the radio-jet AGN feedback can suppress the star formation activity of the surrounding galaxies.   
In order to examine their possible features, we compare the stellar populations of $g$-dropout galaxies existing along the radio major axis with those in the vertical direction for the radio major axis of the faint/luminous HzRGs.

The left (right) panels of Figure \ref{spalign} show the average distributions of the $M_{\mathrm{UV}}$ and the $i-z$ colors of the $g$-dropout galaxies  with $\Delta \phi <30$ degrees and  $\Delta \phi >60$ degrees at a given projected distance from the full (isolated) HzRG sample.  
The error bars of $M_{\mathrm{UV}}$ and $i-z$ in each bin are the standard errors including the photometric and redshift errors. 
We also estimate the averages of $M_{\mathrm{UV}}$ and $i-z$ of the $g$-dropout galaxies in fields, defined by the effective area excluding the regions enclosed by the circles with the 3 arcmin projected radii centered on the full HzRG sample. 
Note that the standard errors of the $g$-dropout galaxies are negligible (about $10^{-4}$) as the number of the $g$-dropout galaxies is critically large.  

Almost no dependency of their $i-z$ colors on the projected distances is found for both of the full and isolated samples.  
On the other hand,  $M_{\mathrm{UV}}$ of the $g$-dropout galaxies existing along the radio major axis tend to be slightly lower than those in  blank fields  at $\lesssim0.2$ pMpc scale with $\gtrsim2\sigma$ significance only for the faint HzRG sample with $L_{1.4\mathrm{GHz}}\sim10^{26.0-26.5}$ W Hz$^{-1}$. 
According to a relation between UV luminosity and star formation rate \citep[][]{Kennicutt98}, the SFRs of the $g$-dropout galaxies with $\Delta\phi<30$ degrees at the projected distance of $\sim0.2$ pMpc are estimated to be $26_{-3}^{+2}$ $M_\odot$ yr$^{-1}$ and  $25_{-3}^{+2}$  $M_\odot$ yr$^{-1}$ for the full and isolated faint samples, respectively, 
while the SFRs of the $g$-dropout galaxies in fields is estimated to be $20$ $M_\odot$ yr$^{-1}$ with an error of about $10^{-3}$ $M_\odot$ yr$^{-1}$. 
This result would support a scenario where an enhanced star formation occurs in the filamentary structures around the HzRGs. 

If large-scale radio-mode AGN feedback is effective, star formation in the surrounding galaxies around HzRGs could be quenched \citep[e.g.,][]{Shabala11}. 
As a result,  $M_{\mathrm{UV}}$ of the $g$-dropout galaxies existing along the radio major axis would be larger than that of fields. 
However, such trend is not found as shown in Figure \ref{spalign}. 
The radio jets might have not yet affected the surrounding galaxies as they have just appeared, or 
the dense gas around the radio galaxy may prevent the jet from spreading far enough \citep[][]{Kawakatu09}. 
Perhaps, the radio-mode feedback feature may be obscured due to the environment study using only star-forming galaxies.
Follow-up observation targeting the possible quenching galaxies is needed to confirm these scenarios.  



\subsection{HzRGs as merger remnants}
If $z\sim4$ HzRGs are merger remnants, they are thought to preferentially appear in the $>4\sigma$ overdense regions where galaxy merger rate is expected to be higher compared to fields \citep[e.g.,][]{Uchiyama20}.  
We, however, found that the proportion of HzRGs associated with the $>4\sigma$ overdense regions is consistent with that of $g$-dropout galaxies within $1\sigma$ error (Section 3.1). 
This result could be explained if the merger rate in such highest overdense regions is relatively lower than that in normal overdense regions, or some HzRGs in the dusty-obscured phase are actually missing even in the deep optical-band surveys down to $i\sim26$ \citep[optically-dark galaxies; ][]{Wang16, Kubo19, Toba20}. 
The former might be explained due to a large velocity dispersion of galaxies in the most overdense regions, which reduce the frequency of mergers \citep[][]{Jian12}. 
The latter is supported by \citet{Kubo19} who suggested that some proportion of optically-dark galaxies hosting active AGNs exist in our protocluster candidate regions. 
\citet{Toba17} also reported that  dust-obscured AGNs tend to show strong clustering signals. 
IR follow-up observations for the protocluster candidates could resolve the underlying physical processes.  

\section{Conclusion} 
We statistically examined the environments of $z\sim4$ HzRGs using the wide and deep imaging data of HSC-SSP. 
The galaxy overdense regions are constructed through the surface density measurement of $g$-dropout galaxies \citep[][]{Toshikawa18}. 
The protocluster candidates are defined by $g$-dropout galaxy overdensities with $>4\sigma$ overdensity significance, which are selected from the effective area  of $\sim121$ deg$^2$ in HSC-Wide regions \citep[][]{Toshikawa18}. 
The HzRG sample is constructed by matching $g$-dropout galaxies to the FIRST radio survey data (Yamashita et al. 2021, in prep.).
The isolated HzRGs that avoid contaminants of low-$z$ are also selected. 
The possible correlation between galaxy overdense regions and properties of HzRGs was examined.
Particularly, we examined if the HzRGs can reside in protocluster candidate regions  and also investigated the dependency of the HzRG density environments on  their radio emission properties such as the radio luminosity and the radio axis orientation. 
The typical halo mass of the HzRGs  was also estimated by clustering analysis. 

Our findings were as follows: 
\begin{itemize}

\item 
No HzRGs reside in the denser regions compared to the $g$-dropout galaxies at log $L_{1.4\text{GHz}} \sim 26.5-27.0$.  
On the other hand, the overdensity significances around faint HzRGs with log $L_{1.4\text{GHz}} \sim 26.0-26.5$ tend to be higher than those around the $g$-dropout galaxies. 
By clustering analysis, the typical HzRG halo mass is estimated to be $\sim10^{13.1} M_\odot$, which is relatively massive compared to $g$-dropout galaxies. 
The difference in halo mass between the faint HzRGs and the $g$-dropout galaxies is found to be larger with higher significance. 
These findings are consistent with the scenario where 
HzRGs get older and more massive as the radio-luminosity decreases. 


\item 
Only one out of the full/isolated HzRG sample is found to be associated with a protocluster candidate within a distance of 1.8 arcmin. 
On the other hand, the HzRG halos are expected to evolve into cluster halos with high probability according to EPS model, 
meaning that non-negligible proportion of HzRGs should be associated with protoclusters.
This disagreement can be explained by taking into account the incompleteness and/or selection bias of the protocluster candidates. 
Spectroscopic follow-up observations for the HzRGs and their surrounding galaxies would lead to the discovery of protoclusters that are missed by overdense region surveys of Lyman break galaxies.  

\item 
The $g$-dropout galaxies within $\lesssim500$ pkpc radius centered on the faint HzRGs 
with $L_{1.4\mathrm{GHz}}\sim10^{26.0-26.5}$ W Hz$^{-1}$ tend to distribute along their radio major axes. 
The UV luminosities of the $g$-dropout galaxies existing along the radio major axises tend to be  higher than those in the field. 
These results imply that the filamentary structures might start to form around the HzRGs.  

\item 
The HzRGs do not preferentially appear in protocluster candidate regions compared to the $g$-dropout galaxies; 
the proportions of HzRGs associated with the protocluster candidates within the 1.8 and 3.0 arcmin angular separation are consistent with those of the $g$-dropout galaxies associated with the protocluster candidates within their separations. 
If the HzRGs are merger remnants, these results imply that the merger rate in the highest overdense regions is relatively lower than that in normal overdense regions, or some HzRGs in the dusty-obscured phase are missing in the overdense regions.

\end{itemize}

We, for the first time, have statistically characterized the HzRG environments at $z\sim4$. 
However, our analysis is based only on photometric information. 
Spectroscopic observations for the HzRGs and their surrounding $g$-dropout galaxies are needed. 
We are going to proceed to analyses at $z<4$ and $z>4$ once the HSC-SSP survey is complete. 
These studies will reveal the co-evolution history of HzRGs and the surrounding galaxies.   


\

\acknowledgments

We are deeply grateful to the referee for his/her helpful comments that improved the manuscript. 
TY acknowledges support from the JSPS grant 21K13968 and 20H01939.
This work is based on data collected at the Subaru Telescope and retrieved from the HSC data archive system, which is operated by Subaru Telescope and Astronomy Data Center at National Astronomical Observatory of Japan. 
The Hyper Suprime-Cam (HSC) collaboration includes the astronomical communities of Japan and Taiwan, and Princeton University. The HSC instrumentation and software were developed by the National Astronomical Observatory of Japan (NAOJ), the Kavli Institute for the Physics and Mathematics of the Universe (Kavli IPMU), the University of Tokyo, the High Energy Accelerator Research Organization (KEK), the Academia Sinica Institute for Astronomy and Astrophysics in Taiwan (ASIAA), and Princeton University. Funding was contributed by the FIRST program from Japanese Cabinet Office, the Ministry of Education, Culture, Sports, Science and Technology (MEXT), the Japan Society for the Promotion of Science (JSPS), Japan Science and Technology Agency (JST), the Toray Science Foundation, NAOJ, Kavli IPMU, KEK, ASIAA, and Princeton University. 

This paper makes use of software developed for the Large Synoptic Survey Telescope. We thank the LSST Project for making their code available as free software at  http://dm.lsst.org

The Pan-STARRS1 Surveys (PS1) have been made possible through contributions of the Institute for Astronomy, the University of Hawaii, the Pan-STARRS Project Office, the Max-Planck Society and its participating institutes, the Max Planck Institute for Astronomy, Heidelberg and the Max Planck Institute for Extraterrestrial Physics, Garching, The Johns Hopkins University, Durham University, the University of Edinburgh, Queen's University Belfast, the Harvard-Smithsonian Center for Astrophysics, the Las Cumbres Observatory Global Telescope Network Incorporated, the National Central University of Taiwan, the Space Telescope Science Institute, the National Aeronautics and Space Administration under Grant No. NNX08AR22G issued through the Planetary Science Division of the NASA Science Mission Directorate, the National Science Foundation under Grant No. AST-1238877, the University of Maryland, and Eotvos Lorand University (ELTE) and the Los Alamos National Laboratory.

\appendix
\section{Calculation of Correlation length, Bias and Halo mass By Cross correlation} 
We describe the detailed calculations of the correlation length, bias, and halo mass of HzRGs through a cross correlation between HzRGs and $g$-dropout galaxies. 

A two-point auto-correlation function (ACF) of $g$-dropout galaxies is given by the following equation \citep[][]{Landy93}: 
\begin{equation}
\omega_{\mathrm{acf}}^{\mathrm{obs}}  (\theta_i) = \frac{DD(\theta_i) - 2 DR(\theta_i) + RR(\theta_i)}{RR(\theta_i)}, 
\end{equation}
where $DD (\theta_i)$, $DR(\theta_i)$, and $RR(\theta_i)$ are the normalized numbers of the galaxy $-$ galaxy pairs,  galaxy $-$ random point pairs, and random point $-$ random point pairs within an angular separation range of [$\theta_i, \theta_i+\delta \theta$) for $i=1,2, \dots, n$, respectively. 
Here, an angular separation space of [1\arcsec, 1000\arcsec) is divided into $n=7$ half-open intervals with a width of $\delta \theta$ in logarithmic space. 
The random points are given by a random catalog of \citet{Coupon18} in which their points are randomly distributed in the survey area with the surface density of $100$ arcmin$^{-2}$. 

A two-point cross-correlation function (CCF) between HzRGs and $g$-dropout galaxies is calculated as follows \citep[e.g.,][]{Wilkinson17}. 
\begin{equation}
\omega_{\mathrm{ccf}}^{\mathrm{obs}}  (\theta_i) = \frac{D_s D_t (\theta_i) - D_s R(\theta_i) - D_t R(\theta_i) + RR(\theta_i)}{RR(\theta_i)}, 
\end{equation}
where $D_s D_t (\theta_i)$,  $D_s R(\theta_i)$ and $D_t R(\theta_i)$ are the normalized numbers of the HzRG $-$ galaxy pairs,  HzRG $-$ random point pairs, and galaxy $-$ random point pairs within a range of an angular separation of [$\theta_i, \theta_i+\delta \theta$), respectively. 

The Jackknife resampling is used to estimate the statistical errors of the ACF and CCF \citep[e.g.,][]{Ishikawa20}. 
The effective survey area is equally divided into $N=26$ subregions. 
The standard deviation of the ACF/CCF for $N$ Jackknife resamplings in a given angular separation of [$\theta_i, \theta_i+\delta \theta$) is given as follows \citep[e.g.,][]{Zehavi05}. 
\begin{equation}
\sigma\left[\omega({\theta_i})\right]= \sqrt{\frac{N-1}{N} \sum\limits_{k=1}^{N} (\omega^k (\theta_i) - \bar{\omega} (\theta_i))^2}, 
\end{equation}
where $\omega^k (\cdot)$ is the ACF/CCF for the $k$ th Jackknife sample, and $\bar{\omega} (\cdot)$ is an average of the ACF/CCF for $N$ Jackknife resamplings.

The true ACF/CCF is obtained by shifting the observed ACF/CCF by an integral constraint \citep[$IC_{\mathrm{acf/ccf}}$; ][]{Groth77}; 
\begin{equation} 
\omega_{\mathrm{acf/ccf}}^{\mathrm{true}} (\theta_i) = \omega_{\mathrm{acf/ccf}}^{\mathrm{obs}} (\theta_i) + IC_{\mathrm{acf/ccf}}, ~~\mathrm{for} ~~ \forall i. \label{eq:10}
\end{equation}
$IC_{\mathrm{acf/ccf}}$ can be calculated by the following equation \citep[e.g.,][]{He18}. 
\begin{equation} 
IC_{\mathrm{acf/ccf}} = \frac{\sum\limits_{i=1}^{n} RR(\theta_i) ~ \omega_{\mathrm{acf/ccf}}^{\mathrm{model}} (\theta_i)}{\sum\limits_{i=1}^{n} RR(\theta_i)}, 
\end{equation}
where $\omega_{\mathrm{acf/ccf}}^{\mathrm{model}} (\cdot)$ expresses a single power law model, 
\begin{equation}
\omega_{\mathrm{acf/ccf}}^{\mathrm{model}} (\theta) \equiv A_\omega \theta^{-\beta} \label{model}
\end{equation}
 with ($A_\omega$,$\beta$) $\in$ ($\mathbb{R}, \mathbb{R}$). 

The spatial correlation length, $r_0$ ($h^{-1}$ Mpc), is estimated through Limber's equation \citep[][]{Limber53}, assuming that the spatial correlation function at $z$, $\xi_{\mathrm{acf/ccf}} (r, z)$, is expressed as $(r/r_0)^{-1-\beta}$ for the ACF/CCF; 
\begin{equation}
r_0 = \left[ \frac{c_0 ~ \omega_{\mathrm{acf/ccf}}^{\mathrm{model}}(\theta)|_{\theta=1 \mathrm{rad}} }{H_0 H_\gamma} \frac{\{ \int dz ~ Q(z) \}^2}{\int dz ~ Q(z)^2 \chi(z)^{-\beta} E(z)} \right]^{\frac{1}{1+\beta}}, \label{eq:corr}
\end{equation}
where $H_\gamma = \Gamma(1/2) \Gamma(\beta/2) / \Gamma((1+\beta)/2)$, $E(z) = \{ \Omega_{\mathrm{m}} (1+z)^3 + \Omega_{\mathrm{\Lambda}}\}^{1/2}$, $\chi(z) = c_0 /H_0 \int_0^z dz \{1/E(z)\}$, and $c_0$ is the speed of light. 
$Q(z)$ is the redshift distribution of the sample, and $\Gamma (\cdot)$ is the Gamma function \citep[e.g.,][]{Toba17, He18}. 
The redshift distribution, $Q(z)$, of the $g$-dropout galaxies has been already estimated 
using the same HSC catalog and sample selection criteria as this work by \citet{Ono18}. 
We assume that the redshift distribution of the HzRGs are the same as that of the $g$-dropout galaxies,  because the HSC counterparts of the HzRGs are identically selected to the $g$-dropout galaxies. 
Then, the correlation lengths are estimated  through the same equation (\ref{eq:corr}) for both of the ACF and CCF \citep[e.g.,][]{Croom99}.

The spatial correlation function, $\xi_{\mathrm{acf/ccf}} (r, z)$, for the ACF/CCF is associated with that of the underlying dark matter halo, $\xi_{\mathrm{dm}} (r, z)$, via the bias factor $b_{\mathrm{acf/ccf}}(z)$; 
\begin{equation} 
\xi_{\mathrm{acf/ccf}} (r, z) |_{r=8 h^{-1} \mathrm{Mpc}} = b_{\mathrm{acf/ccf}}(z)^2 \xi_{\mathrm{dm}} (r,z)|_{r=8 h^{-1} \mathrm{Mpc}}. 
\end{equation}
The $\xi_{\mathrm{dm}} (8,z) (\equiv \xi_{\mathrm{dm}} (r,z)|_{r=8 h^{-1} \mathrm{Mpc}})$ can be estimated from the cold dark matter model \citep[][]{Myers06} through the following equations.  
\begin{eqnarray}
\xi_{\mathrm{dm}} (8,z) &=& \frac{(2-\beta)(3-\beta)(5-\beta)2^{1+\beta}}{72} \left[ \sigma_8 \frac{g(z)}{g(0)} \frac{1}{z+1} \right]^2, \\
g(z) &=&\frac{5\Omega_{\mathrm{m}} (z)}{2} \left[ \Omega_{\mathrm{m}} (z)^{4/7} -  \Omega_{\mathrm{\Lambda}} (z) + \left(1+\frac{\Omega_{\mathrm{m}} (z)}{2} \right) \left(1+ \frac{\Omega_{\mathrm{\Lambda}} (z)}{70} \right) \right]^{-1}, \\
\Omega_{\mathrm{m}} (z) &=& \frac{\Omega_{\mathrm{m}} (1+z)^3}{E(z)^2}, \\
\Omega_{\mathrm{\Lambda}} (z) &=&  \frac{\Omega_{\mathrm{\Lambda}}}{E(z)^2}. 
\end{eqnarray}
The bias factors of the HzRGs, $b_{\mathrm{hzrg}}(z)$, and the $g$-dropout galaxies, $b_{\mathrm{g}} (z)$, are expressed as  
\begin{eqnarray} 
b_{\mathrm{hzrg}} (z) &\equiv& \frac{b_{\mathrm{ccf}} (z)^2}{b_{\mathrm{acf}} (z) }, \\
b_{\mathrm{g}}   (z)    &\equiv& b_{\mathrm{acf}} (z) . 
\end{eqnarray}

The typical underlying dark matter halo mass of the HzRGs/$g$-dropout galaxies, $M_{\mathrm{h}}^{\mathrm{hzrg/g}}$, can be expressed as an inverse function of its bias factor, according to $N$-body simulation of \citet{Sheth01}; 
\begin{eqnarray}
b_{\mathrm{hzrg/g}} (z) &=& 1+\frac{1}{\sqrt{a}\delta_\mathrm{c} (z)} \left[ \sqrt{a} (a\nu^2) + \sqrt{a} b (a\nu^2)^{1-c}  - \frac{(a\nu^2)^c}{(a\nu^2)^c + b (1-c) (1-c/2)} \right], \label{aa}
\end{eqnarray}
where 
\begin{eqnarray}
\nu &=& \frac{\delta_c}{\sigma(M_{\mathrm{h}}^{\mathrm{hzrg/g}}) D(z)} \\ 
\sigma(M_{\mathrm{h}}^{\mathrm{hzrg/g}}) & = & \sqrt{\int dk \frac{\Delta(k)^2 W(kR)^2}{k}} \\ 
W(kR) &=& \frac{3 \sin(kR) - kR \cos(kR)}{(kR)^3} \\
R &=& \left( \frac{3 M_{\mathrm{h}}^{\mathrm{hzrg/g}}}{4 \pi \rho_0} \right)^{\frac{1}{3}}. \label{bb}
\end{eqnarray}
The critical density, $\delta_c$, is $1.686$ and $(a, b, c) =( 0.707, 0.5, 0.6)$ \citep[][]{Tinker10}. 
The growth factor, $D(z)$, is approximately equal to $g(z)/(1+z)$ \citep[][]{Carroll92}. 
The linear power spectrum, $P\equiv \Delta(k)^2$, is calculated by the {\tt CAMB} package \citep[][]{Lewis00, Challinor11}.

\end{document}